\documentclass[twocolumn,showpacs,preprintnumbers,amsmath,amssymb,superscriptaddress]{revtex4}

\usepackage{graphicx}
\usepackage{amsmath}
\usepackage{epsfig}
\usepackage{bm}

\def\>{{\rangle}}
\def\<{{\langle}}
\newcommand\Tr{\mathop\mathrm{Tr}\nolimits}

\begin{document}
   
\title{Nonequilibrium Transport of  Quantum Molecular Chain in terms of the Complex Liouvillian Spectrum} 

\author{Satoshi \surname{Tanaka}}
\email{stanaka@p.s.cias.osakafu-u.ac.jp}
\affiliation{Department of Physical Science, Osaka Prefecture 
University, Gakuen-cho 1-1, Sakai 599-8531, Japan}
\author{Kazuki \surname{Kanki}}
\affiliation{Department of Physical Science, Osaka Prefecture 
University, Gakuen-cho 1-1, Sakai 599-8531, Japan}
\author{Tomio Petrosky}
\affiliation{Center for Studies in Statistical Mechanics and Complex
Systems, The University of Texas at Austin, Austin, TX 78712 USA} 

\date{\today}

\begin{abstract}
Transport process in molecular chain in nonequilibrium stationary state is theoretically investigated.
The molecule is interacting at its both ends with thermal baths which has different temperatures, while no dissipation mechanism is contained inside the molecular chain.
We have first obtained the nonequilibrium stationary state outside the Hilbert space in terms of the complex spectral representation of Liouvillian.
The nonequilibrium stationary state is obtained as an eigenstate of the Liouvillian which is constructed through  the collision invariant of the kinetic equation.
The eigenstate of the Liouvillian contains an information of spatial correlation between the molecular chain and the thermal baths.
While energy flow in the nonequilibrium state which is due to the first order correlation can be described by Landauer formula, the particle current due to the second order correlation cannot be described by the Landauer formula.
The present method provides a simple and perspective way to evaluate the energy transport of molecular chain under the nonequilibrium situation.
\end{abstract}

\maketitle
\vfill

\section{Introduction}


Recently there have been paid much attention  to nonequilibrium transport processes through molecular wire junctions.\cite{JortnerBook,Cahill03}
As the system is shrunk down to a smaller size than the mean free path of a carrier, such as an electron, a phonon, or an exciton, the ballistic quantum transport exhibits a characteristically different feature from the corresponding bulk properties.
For example, it was discovered that the electronic or thermal conductance of sub-micron size one-dimensional chain  is quantized at low temperature.\cite{Rego98,Schwab00,Wees88,Wharam88}
When the system size is further shrunk down to the nanometer size, a discretized energy level structure of a nanowire plays an important role in transport processes through the resonance effect.
Recent ultrafast nonlinear optical spectroscopy has revealed the temporal behavior of relaxation process of the photo-excited molecular chain.\cite{Schwarzer04,Wang07}
Also in biomolecules, such as alpha-helix protein and DNA strands, which consists of  molecular chain structures, the transport process in nonequilibrium state have been extensively studied in order to clarify the biofunctions of these molecules from  a microscopic view point.\cite{Berlin02,Botan07,Nguyen10,Kawai10,Tanaka09} 
 
In order to understand the transport processes in the nonequilibrium stationary state, we need to know first of all how to describe the nonequilibrium stationary state with which a physical quantity can be obtained as an expectation value of an observable.
While it is obvious that a stationary state is described by a canonical distribution in a thermal equilibrium, it is not so easy to find the explicit form of the stationary state for  nonequilibrium situation.
Meanwhile a phenomenological Landauer formula has been extensively used as a useful tool to evaluate a transport coefficient in a nonequilibrium stationary state without explicitly describing the stationary state.
There have been enormous works trying to make clear its microscopic foundation and its applicability in terms of nonequilibrium Green's function method, quantum Langevin method, and so on.
\cite{JortnerBook,Landauer57,Rieder67,Zurcher90,Segal01,Segal03,Petrov05,Galperin07,Subotnik09}

On the other hand, Prigogine and his coworkers have developed a theory of the nonequilibrium statistical mechanics in terms of the complex spectral representation of the Liouville-von Neumann super-operator, or simply called {\it Liouvillian}.\cite{PrigogineText,Petrosky99}
They have clarified that the eigenvalue problem of the Liouvillian is classified into independent subspaces according to the order of correlations based on the concept of the {\it dynamics of the correlation} where the fundamental object is a correlation.
In a classical gas system, for example, each correlation is characterized by the dependence of the distribution function on the wave number which characterizes the spatial correlation: The inhomogeneity is a correlation component which has a single non-vanishing wave number of a particle, while {\it inter-particle correlation} is a component which has several non-vanishing wave numbers of different particles.\cite{Petrosky97,Petrosky02,Petrosky10} 

This classification in terms of the correlation enables to transform the eigenvalue problem of the Liouvillian to the eigenvalue problem of the collision operator of the kinetic equation in a correlation subspace.
It is well known that the collision operator plays a central role in nonequilibrium statistical mechanics, as seen in Boltzmann equation or Fokker-Planck equation.\cite{PrigogineText,ResiboisBook}
A striking finding of the theory is that the spectrum of the collision operator is identical with that of the Liouvillian, which signifies the direct link between the microscopic dynamics governed by the Liouvillian and the phenomenological kinetic theory.\cite{Petrosky96,Petrosky10,Petrosky97}
Since the collision operator is non-Hermite operator, the eigenvalues can take complex values which reflects the dissipation of the system.
The eigenstate of the Liouvillian is obtained from the eigenstate of the collision operator by operating a creation-of-correlation operator onto the eigenstate of the collision operator which represents the transition from a privileged correlation subspace to other non-privileged correlation subspaces.\cite{Petrosky96,Petrosky97}
The eigenstate of the Liouvillian is then represented as a series of the correlation generated by the dynamics of correlation.


There have been a few applications of the theory to real physical systems,\cite{Petrosky99,Petrosky02,Tay06} but the transport processes in nonequilibrium situations have not been fully investigated yet in the context of dynamics of correlation.
Our aim here is to apply the theory to the real mesoscopic system under nonequilibrium situation and to systematically derive the transport quantities in a stationary state.

In the present work, we consider a molecular chain coupled with different thermal baths at the both ends, where a quantum particle is confined in the molecule and it ballistically transfers within the molecule.
An exchange of energy occurs between the molecule and the thermal baths, while there is no particle exchange between them as shown in Fig.\ref{Fig1}.
The nonequilibrium stationary state is obtained as the zero eigenstate of the Liouvillian which is represented by superposition of different order of correlations.
A leading term is the non-correlation component, or called {\it vacuum of correlation}, followed by the higher order correlation components.
As will be shown, the inhomogeneity component of a single particle is created by the second order interaction from the vacuum of correlation.
This is contrast to the case of a well-known gas system where the inhomogeneity component is disconnected from the vacuum of correlation subspace by the interaction, thus is not created from the vacuum of correlation.\cite{PrigogineText}

Once the nonequilibrium stationary state is obtained as a zero eigenstate of the Liouvillian, a physical quantity is evaluated as an expectation value of a corresponding observable in the stationary state.
Since the nonequilibrium stationary state is an eigenstate of the Liouvillian, the expression of a physical quantity is justified from the microscopic dynamics without any phenomenological assumptions.

We have derived an expression of the energy flow with use of the nonequilibrium stationary state, and found that the first order correlation is responsible for the energy flow.
We shall show that the expression of the energy flow is cast into the Landauer formula with a characteristic transmission function,
because the first oder correlation is directly related to a collision operator which gives a transition probability between the molecular states, represented by Fermi's golden rule. 

On the other hand, the physical quantity described by the higher order correlation cannot be reduced in the Landauer formula.
As an example, we consider the particle current which is induced by external thermal force,  as in the mechanical force when a molecule is subjected to an external electric field.
However, it is well-known that the treatment of the thermal force is in general much more complex than a mechanical force, because the thermal force stems from a many-body dissipative effect.\cite{BalescuBook}
We shall reveal that the particle current is attributed to the second order correlation component.
Since there is no direct connection between the second order correlation and the collision operator, 
 we cannot cast it into the Landauer formula unlike the case of the energy flow in this problem.
 
In Section \ref{Sec:Model} we present a model Hamiltonian of the molecular chain coupled with different thermal baths at its both ends.
The eigenvalue problem of the Liouvillian is presented in Section \ref{Sec:Eigen} to obtain the nonequilibrium stationary state as zero eigenstate of the Liouvillian.
Energy flow is obtained with use of the nonequilibrium stationary state and Landauer formula is derived in Section \ref{Sec:EFlow} where we explain why the energy flow can be cast into the Landauer formula.
In Section \ref{Sec:Pflow}, the induced polarization, or its conjugate current, is evaluated as an example of the higher order correlation which cannot be reduced to Landauer formula.
We illustrate in Section \ref {Sec:Discussion} an application to DNA molecular chain.
In Section \ref{Sec:Remarks} we give some concluding remarks.
 
For readers who are not familiar with the complex spectral representation of the Liouvillian in Appendix \ref {App:LiouvilleSpace} we introduce the Liouville space and show the explicit expression of the interaction in terms of the Liouville basis, respectively.
The complex spectral representation of Liouvillian is summarized in Appendix \ref {App:Complex} where the eigenstate of the Liouvillian is given as a functional of the eigenstate of the collision operator.
With use of the explicit expression of the interaction, we derive the collision operator in Appendix \ref {App:DerivKinEq}.
In Appendix \ref {App:BathChange} we derive a formula which we use in Section \ref {Sec:Eigen}.

 \section{Model}\label{Sec:Model}
 
\begin{figure}
\begin{center}
\includegraphics[width=8cm]{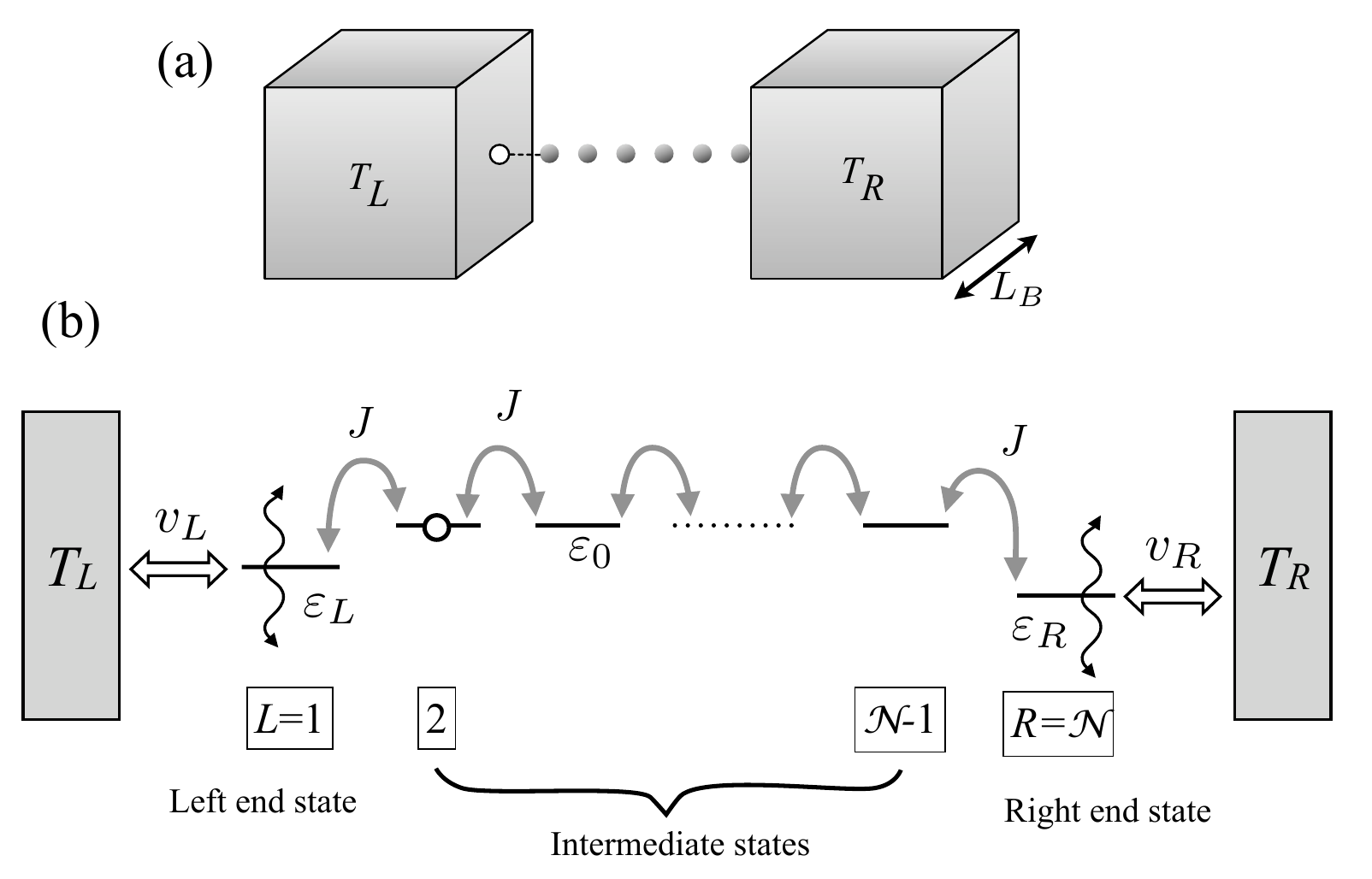}
\caption{One-dimensional molecular chain composed of ${\cal N}$ molecular units each of which possesses a bound state.
The left- and right-end states have the energies of $\varepsilon_{1}=\varepsilon_{L}$ and $\varepsilon_{\cal N}=\varepsilon_{R}$, respectively, while 
the other states have the same energy of $\varepsilon_{m}=\varepsilon_{0}$ for $m=2,\dots, {\cal N}-1$.
A particle transfers between these states with the transfer integral of $J$.
The molecular chain interacts with the two thermal baths with different temperatures of $T_{L}$ and $T_{R}$ at the both ends of the molecule.
 }
\label{Fig1} 
\end{center}
\end{figure}

We consider a one-dimensional molecular chain consisting in $\cal N$-molecular units each of which contains a single bound state.
This molecular chain is coupled with two thermal baths at the both ends.
The total Hamiltonian is written as
\begin{equation}
\label{Htot}
 H=H_M+H_{B}+g H_{MB} \;,
 \end{equation}
where $H_{M}$ and $H_{B}$ describe the molecular chain and the thermal baths, respectively, and the interaction is represented by $H_{MB}$ with a dimensionless coupling constant $g$.

We consider that a quantum particle transfers in the molecular chain.
The molecular Hamiltonian is then described by a one-dimensional tight binding Hamiltonian:
\begin{equation}
\label{HM}
H_{M}=  \sum_{m=1}^{\cal N} \varepsilon_{m}|m\>\<m|  
-  \sum_{\<m,m'\>} J_{m,m'} |m\>\<m'| , 
\end{equation}
where $\varepsilon_{m}$ is an energy of the bound state $|m\>$ at the $m$-th molecular unit as shown in Fig.\ref{Fig1}.
We denote the left end state as $|L\>\equiv |1\>$ and the right end state as $|R\>\equiv|N\>$. 
We assume that $\varepsilon_{m}=\varepsilon_{0}$ in the middle of the chain for $m=2,\dots,{\cal N}-1$, while the left and the right end states have different energies of $\varepsilon_{L}$ and $\varepsilon_{R}$.
The second term represents the particle transfer where we take into account only the nearest neighbor transfer; $\<m,m'\>$ in the second term denotes taking a sum of the nearest neighbor bound states.
In the present work, we assume a constant value for the transfer integrals of $J_{m,m'}=J$ for any $m$ and $m'$, and we take  $J=1$ as an energy unit which also becomes a unit of temperature. 

The eigenvalue problem of the molecular Hamiltonian $H_{M}$ with the $\cal N$ dimension is solved to obtain the $\cal N$ eigenstates as
\begin{equation}
\label{HmEVP}
H_M| E_{\bar j}\>=E_{\bar j}|E_{\bar j}\>   \quad ( \bar j=1,\cdots,{\cal N})  \;,
\end{equation}
where an eigenstate $|E_{\bar j}\>$ is represented by 
\begin{equation}
\label{Barj}
|E_{\bar{j}}\>=\sum_{m=1}^{\cal N} c_{m,\bar j} |m\> \;.
\end{equation}
Hereafter in order to avoid a heavy notation, we simply describe $|E_{\bar j}\>$ as $|\bar j\>$. 

Both of the left and right thermal baths are assumed to be a three-dimensional harmonic crystal described by a Debye model in a large box of  volume $L_{B}^{3}$. 
The Hamiltonian of the thermal bath systems reads
\begin{eqnarray}
\label{HB}
H_B=\sum_{r=L,R}\sum_{\bf q}\hbar\omega_{r,\bf q}b^\dagger_{r,\bf q}b_{r,\bf q} \;,
\end{eqnarray}
where $b_{r,{\bf q}}$ $(r=L,R)$ are the annihilation operators of the phonons of the thermal baths with the energy dispersion given by $\omega_{r,{\bf q}}=c|{\bf q}|$.
We take the box normalization with a periodic boundary condition for the thermal bath systems which gives the discrete wave vectors for the phonon mode as
\begin{equation}
{\bf q}_{j}={\bf j}\triangle q \;,
\end{equation}
where ${\bf j}$ is a three-dimensional integer vector and $\triangle q\equiv 2\pi/L_{B}$.
In the large volume limit for the bath systems ($L_{B}\to\infty$), we have  
\begin{equation}\label{LargeLim}
{1\over \Omega}\sum_{{\bf q}_{j}}\cdots \longrightarrow \int d{\bf q} \cdots  \;, \; \Omega \delta_{{\bf q},0} \longrightarrow \delta({\bf q}) \;,
\end{equation}
where  $\Omega\equiv (1/\triangle q)^{3}$.

With use of the density of states of the thermal phonon system per volume ${\cal D}_{ph}(\omega)$, we can change the integral of $\bf q$ to the integral of $\omega$ as 
\begin{equation}\label{IntDOS}
{1\over \Omega}\sum_{{\bf q}_{j}} \xrightarrow{\Omega\to\infty}  \int d{\bf q} \cdots =\int_{0}^{\infty} d\omega {\cal D}_{ph}(\omega) \cdots \;,
\end{equation}
where ${\cal D}_{ph}(\omega)$ is given by 
\begin{eqnarray}\label{phonon DOS}
{\cal D}_{ph}(\omega)=
\begin{cases}
{ 4\pi  \omega^{2}\over c^{3} }    & \mbox{for } 0\le \omega \le \omega_{D} \\
0 \;  &\mbox{otherwise} \;.
\end{cases}
\end{eqnarray}
In Eq.(\ref{phonon DOS}) Debye frequency $\omega_{D}$ is given by
\begin{equation}
\omega_{D}=c\left( {3 N_{B} \over 4\pi \Omega} \right)^{1/3} \;, 
\end{equation}
where $N_{B}$ is a number of the normal phonon modes.
 In the present work, we take $\omega_{D}$ to be large enough compared to the other parameters so that the value of $\omega_{D}$ will not affect the results. 

The unperturbed Hamiltonian $H_{0}$ is then defined as a sum of $H_{M}$ and $H_{B}$:
\begin{equation}\label{H0}
H_{0}=H_{M}+H_{B}\;.
\end{equation}

\begin{figure}
\begin{center}
\includegraphics[width=8cm]{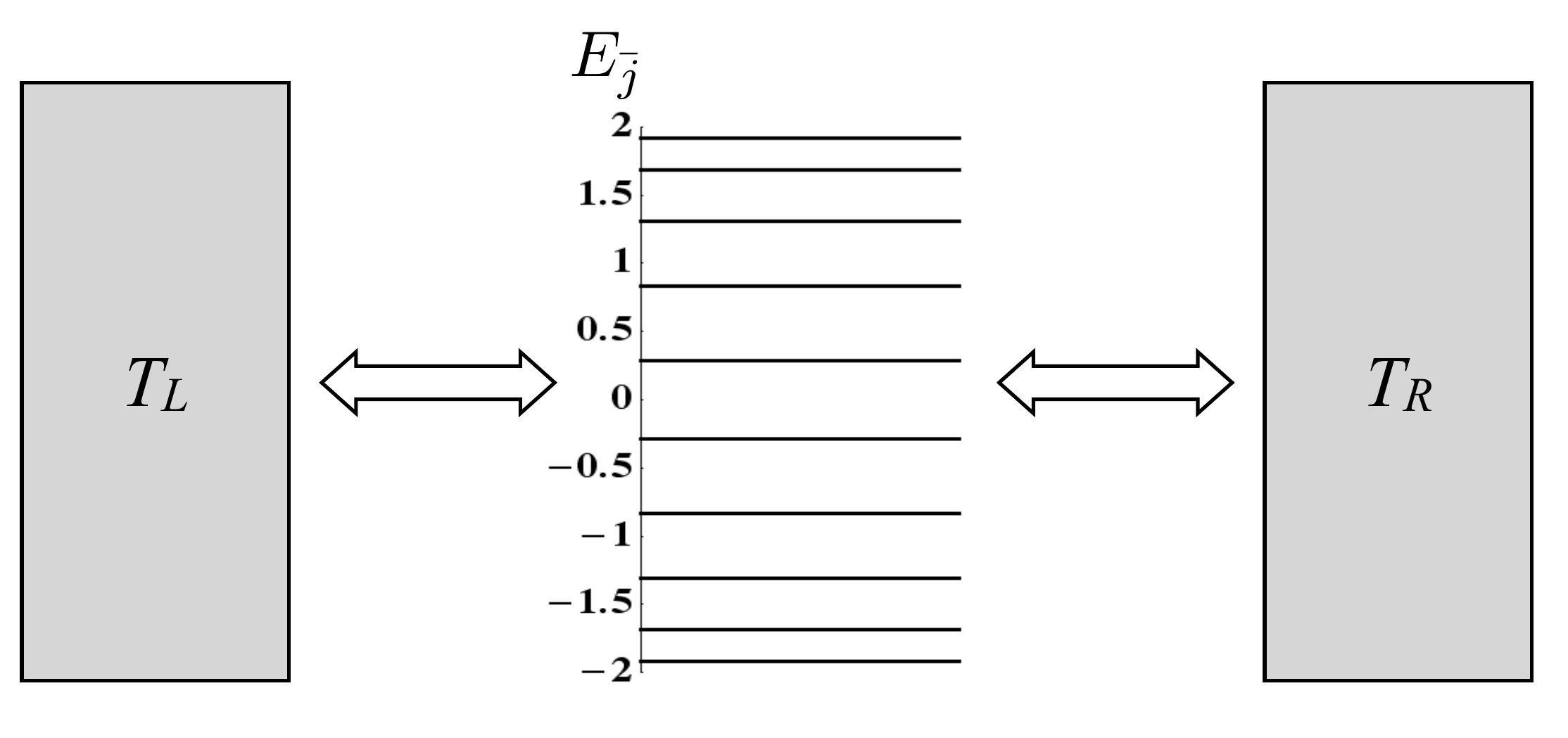}
\caption{Molecular level and the interaction scheme.
 The molecular eigenstates $|\bar j\>$ are coupled with the thermal baths.}
\label{Fig:Scheme} 
\end{center}
\end{figure}

We consider the interaction between the molecule and the thermal baths represented by
\begin{eqnarray}
\label{HMB}
g H_{MB}&=&{g\over \sqrt{\Omega}} \sum_{\bf q} v_{L,{\bf q}} |L\>\<L|( b_{L,{\bf q}}+b_{L,{\bf q} }^{\dagger}) \nonumber\\
&&+ {g\over \sqrt{\Omega}}\sum_{\bf q} v_{R,{\bf q}} |R\>\<R|( b_{R,{\bf q}}+b_{R,{\bf q}}^{\dagger})  \nonumber \\
&&\equiv g ( H_{MB}^{L}+H_{MB}^{R}  ) \;,
\end{eqnarray}
where $v_{r,{\bf q}}$ $(r=L,R)$ are the interaction potentials, and we assume that  $v_{r,{\bf q}}$ is independent of  $\Omega$ and ${\bf q}$, i.e.,  $v_{r,{\bf q}}=v_{r}$.
We consider that in Eq.(\ref{HMB}) the molecular chain is interacting with a thermal bath only at the both ends of the chain so that the energies of the end states $|L\>$ and $|R\>$ are fluctuated by the interaction.
With use of the eigenstates of $H_{M}$, the interaction Hamiltonian $ H_{MB}$ is represented by
\begin{equation}
\<\bar j'|g H_{MB}|\bar j\>={g \over \sqrt{\Omega}}\sum_{r=L,R}\sum_{\bf q}  v_r c_{r,\bar j'} c_{r,\bar j}(b_{r,{\bf q}}+b_{r,{\bf q}}^{\dagger} ) \;.
\end{equation}
In Fig.\ref{Fig:Scheme}, we draw the molecular eigenstates $|\bar j\>$ of $H_{M}$ with the energy $E_{\bar j}$.
These states are coupled with thermal baths through the coupling of the end state components.

\section{Nonequilibrium state as an eigenstate of Liouvillian}\label{Sec:Eigen}

The time evolution of  the total system obeys the Liouville-von Neumann equation
\begin{equation}
i \frac{\partial}{\partial t} \rho (t)    =\mathcal{L}\rho(t)  \;,
\label{Liouville}
\end{equation}
where $\rho(t)$ is a density matrix of the total system and  $\mathcal{L}$ is the Liouvillian  defined by
$
\mathcal{L}\rho \equiv  [H,\rho]/\hbar .
$
We assume that in the initial state the left and right thermal baths are in thermal equilibrium with different temperatures $T_{L}$ and $T_{R}$, respectively:
\begin{eqnarray}
\label{EqBath}
\rho_{r}^{eq}=\frac{\exp\big[-\sum_q \hbar\omega_{r,q}b^\dagger_{r,q}b_{r,q} / k_{B}T_{r}\big]}{Z_{r}}   \quad (r=L,R) \;,
\end{eqnarray}
where
$
Z_{r} =\prod_q \big(1-\exp[ -\hbar\omega_{r,q}/k_{B}T_{r}]\big)^{-1} 
$  with  $k_{B}$ the Boltzmann constant.
Hereafter we take $k_{B}=1$.
The thermal distribution $\rho_{r}^{eq}$ gives the Planck's distribution
\begin{equation}\label{Planck}
n_{r}(\omega)={1\over \exp[-\hbar\omega/T_{r}]-1 } \;.
\end{equation}

In the present work, we shall use the Liouville space representation which is briefly summarized in Appendix \ref {App:LiouvilleSpace}.
As a basis set for the particle system we shall  introduce a Wigner representation defined by
\begin{equation}\label{WignerPtcl}
|{\eta},{Y}\>\!\>\equiv|Y+{\eta\over 2};Y-{\eta \over 2}\>\!\>\equiv |Y+{\eta\over 2}\>\<Y-{\eta \over 2}|  \;,
\end{equation}
where $|Y+{\eta / 2}\>$ and $\<Y-{\eta  / 2}| $ are the eigenstates of  $H_{M}$: $|\bar j\>=|Y+\eta/ 2\>$ and $|\bar j'\>=|Y-\eta/ 2\>$.   
Then $|\eta,Y\>\!\>$ becomes an eigenstate of ${\cal L}_{M}$:
\begin{equation}\label{LMEV}
{\cal L}_{M}|\eta, Y\>\!\>=\Delta_{\eta,Y}|\eta,Y\>\!\>
\end{equation}
with the eigenvalue of 
\begin{equation}\label{DelDef}
\Delta_{\eta,Y}\equiv {1\over \hbar}\left( E_{Y+{\eta\over2}}-E_{Y-{\eta\over 2}} \right)  \;.
\end{equation}
Note the property 
\begin{equation}\label{PropDelta}
\Delta_{-\eta,Y}=-\Delta_{\eta,Y}   \;.
\end{equation}
By this definition, $\eta=0$ and $\eta\neq 0$ represent the diagonal and the off-diagonal components of the density matrix of the particle, respectively.
Note that the representation of $|\eta,Y\>\!\>$ corresponds to the Wigner basis representation of $|k,P\>\!\>$ of a Boltzmann gas system where $k$ and $P$ correspond to $\eta$ and $Y$, respectively.

Similarly the Wigner representation is defined for the phonon systems, we write the usual Wigner representation for a  $(r,{\bf q})$ phonon mode as
\begin{equation}\label{WignerBath}
|\nu_{r,{\bf q}}, N_{r,{\bf q}}\>\!\>\equiv |n_{r,{\bf q}};n'_{r,{\bf q}} \>\!\> \;,
\end{equation}
where
\begin{equation}
\nu_{r,{\bf q}}\equiv n_{r,{\bf q}}-n'_{r,{\bf q}} \;, \; N_{r,{\bf q}}\equiv { n_{r,{\bf q}}+n'_{r,{\bf q}} \over 2 } \;.
\end{equation}
The Wigner basis $|\{\nu \},\{N\} \>\!\>$ satisfies 
\begin{equation}
{\cal L}_{B}|\{\nu \},\{N\} \>\!\> =\nu\omega |\{\nu \},\{N\} \>\!\> \;,
\end{equation}
where $\{\cdots\}$ denotes a set of all the phonon normal modes and $\nu\omega\equiv \sum_{r=L,R}\sum_{\bf q}\nu_{r,{\bf q}}\omega_{r,{\bf q}}$.
 
The eigenstates of the unperturbed Liouvillian is then represented by a tensor products of $|\eta, Y\>\!\>$ and $|\{\nu\},\{N\}\>\!\>$ as
\begin{equation}\label{L0EV}
{\cal L}_{0}|\eta, Y\>\!\>\!\otimes\! |\{\nu\},\{N\}\>\!\>=\left( \Delta_{\eta,Y}+\nu\omega \right) |\eta, Y\>\!\>\!\otimes\! |\{\nu\},\{N\}\>\!\> \;.
\end{equation}

In terms of the Wigner basis of the eigenstates of $\mathcal{L}_{0}$, we can classify the Liouville space according to the order of correlations.
For that purpose, we introduce the projection operators that specify the correlation components, such as a one-particle distribution of the particle $\hat{\cal P}^{(\eta)}$, correlation between the particle and the phonon systems $\hat{\cal P}^{(\eta,\nu_{r,\bf q})}$, and so on:
\begin{subequations} \label{Pclass}
\begin{eqnarray}
&& \hat{\cal P}^{(\eta)}\equiv\sum_{\{N\}}\sum_{Y}  |\eta, Y\>\!\>{\<\!\<}\eta,Y| 
 \otimes |\{0\},\{N\} \>\!\>\<\!\<\{ 0\}, \{N\}|  \; , \nonumber\\
 && \\
&&\hat{\cal P}^{(\eta,\nu_{r,{\bf q}})}=\sum_{\{N\}}\sum_{Y}|\eta,Y\>\!\>{\<\!\<}\eta,Y|\nonumber\\
&&\otimes |\nu_{r,{\bf q}},\{0\}'_{\bf q},N_{r,{\bf q}},\{N\}_{\bf q}' \>\!\>\<\!\<\nu_{r,{\bf q}},\{0\}'_{\bf q},N_{r,{\bf q}},\{N\}'_{\bf q}|  \; ,  \nonumber\\
&& \\
&&\hat{\cal P}^{(\eta,\nu_{r,{\bf q}_{1}},\nu_{r,{\bf q}_{2}})}=\sum_{\{N\}}\sum_{Y}|\eta,Y\>\!\>{\<\!\<}\eta,Y|\nonumber\\
&&\otimes |\nu_{r,{\bf q}_{1}},\nu_{r,{\bf q}_{2}},\{0\}'_{{\bf q}_{1},{\bf q}_{2}},N_{r,{\bf q_{1}}},N_{r,{\bf q_{2}}},\{N\}'_{{\bf q}_{1},{\bf q}_{2}} \>\!\> \nonumber\\
&&\qquad \<\!\<\nu_{r,{\bf q}_{1}},\nu_{r,{\bf q}_{2}},\{0\}'_{{\bf q}_{1},{\bf q}_{2}},N_{r,{\bf q_{1}}},N_{r,{\bf q_{2}}},\{N\}'_{{\bf q}_{1},{\bf q}_{2}}|  \; ,  \nonumber\\
&& \qquad \dots \;,
\end{eqnarray}
\end{subequations}
where $\{\cdots\}'_{{\bf q}_{1},{\bf q}_{2},\cdots}$ means a set of all the normal modes other than ${\bf q}_{1},{\bf q}_{2},\cdots$.
In order to consider the nonequilibrium stationary state which is obtained in the long time limit,  it is appropriate to consider
eigenvalue problem of ${\cal L}$, instead of solving Eq.(\ref{Liouville}) as an initial value problem.
For the system coupled with the thermal baths with an infinite degrees of freedoms, $\cal L$ may have a complex spectrum.
As shown in Appendix \ref{App:Complex}, the complex eigenvalue problem of the Liouvillian is classified by the correlations:
\begin{equation}
\label{LEVProb}
\mathcal{L}|F_\alpha^{(\mu)})\!)=Z_\alpha^{(\mu)}|F_\alpha^{(\mu)})\!) \; , \;  (\!(\tilde{F}_\alpha^{(\mu)}|\mathcal{L}=(\!(\tilde{F}_\alpha^{(\mu)}|Z_\alpha^{(\mu)} \;,
\end{equation}
where $(\mu)$ is a combined index of $(\eta, \{\nu\})$ in Eqs.(\ref{Pclass}) and $\alpha$ is an index of an eigenstate of a $(\mu)$-subspace. 
The eigenstates of  $|F_\alpha^{(\mu)})\!)$ and $(\!(\tilde{F}_\alpha^{(\mu)}|$ are the left- and right-eigenstates of ${\cal L}$ with the complex eigenvalues $Z_\alpha^{(\mu)}$ for the total system composed of the particle and phonon  systems.
In Eq.(\ref{LEVProb}), we have used $|\cdot )\!)$ notation instead of $|\cdot\>\!\>$ for the total system consisting in the particle and phonon systems.
We briefly summarize the complex spectral representation of Liouvillian in Appendix \ref {App:Complex}, and the reader could consult Refs.\cite{Petrosky97,Petrosky99,Petrosky10} for the details.
Since the nonequilibrium stationary state is time independent and is achieved in the long time, we should seek for the zero eigenstate of ${\cal L}$.   \par

As have shown in Appendix \ref{App:Complex}, by acting the projection operators on Eq.(\ref {LEVProb}), we have the eigenvalue problem of a collision operator given by
\begin{equation}\label{App:EVPCol}
 \hat{\Psi}^{(\mu)} (Z_j^{(\mu)})|u_j^{(\mu)})\!) = Z_j^{(\mu)} | u_j^{(\mu)} )\!)   \; ,
\end{equation}
where 
\begin{equation}\label{App:uj}
 |u_j^{(\mu)} )\!) = (N_j^{(\mu)})^{-1/2} \hat{\cal P}^{(\mu)} |F_j^{(\mu)} )\!) 
\end{equation}
is a {\it privileged} component of $|F_{j}^{(\mu)})\!)$ and $N_j^{(\mu)}$ is a normalization constant which is given by Eq.(\ref{App:Nnorm}).  \par

Here, $\hat{\Psi}^{(\mu)}$ is the {\it collision operator} familiar to nonequilibrium statistical mechanics.\cite{PrigogineText, Resibois67, ResiboisBook,BalescuBook} 
This operator is associated to {\it diagonal transitions} between two states corresponding to the same projection operator $\hat{\cal P}^{(\mu)}$:
\begin{equation}\label{App:Psi}
 \hat{\Psi}^{(\mu)}(z)={\hat{\cal P}}^{(\mu)} \mathcal{L}_0 \hat{\cal P}^{(\mu)} 
+\hat{\cal P}^{(\mu)}g\mathcal{L}_{MB}\hat{\cal Q}^{(\mu)}\hat{\mathcal{C}}^{(\mu)}(z) \hat{\cal P}^{(\mu)} \; ,
\end{equation}
where $\hat{\cal Q}^{(\mu)}\equiv 1-\hat{\cal P}^{(\mu)}$.
In Eq.(\ref{App:Psi})
\begin{equation}\label{App:Creation}
 \hat{\mathcal{C}}^{(\mu)}(z)=\frac{1}{z-\hat{\cal Q}^{(\mu)}\mathcal{L}\hat{\cal Q}^{(\mu)}}\hat{\cal Q}^{(\mu)}g\mathcal{L}_{MB}\hat{\cal P}^{(\mu)}
\end{equation}
is called the creation-of-correlation operator, or simply the {\it creation operator}.\cite{Petrosky97}
Eqs.(\ref{LEVProb}) and (\ref{App:EVPCol}) show that the spectra of $\cal L$ and $\hat\Psi$ are identical.
This directly links the microscopic dynamics to macroscopic kinetic processes. \par

We can expand the creation operator in a series expansion of $g$ as
\begin{eqnarray}\label{Cexpand}
 \hat{\mathcal{C}}^{(\mu)}(z)&=&{1 \over z-\hat{\cal Q}^{(\mu)}\mathcal{L}_0 \hat{\cal Q}^{(\mu)}} \nonumber\\
&&\times  \sum_{\xi=0}^\infty  g^{\xi}
\left ( \hat{\cal Q}^{(\mu)} \mathcal{L}_{MB} \hat{\cal Q}^{(\mu)} 
 {1 \over z-\hat{\cal Q}^{(\mu)}\mathcal{L}_0 \hat{\cal Q}^{(\mu)}} \right)^\xi \nonumber\\
 &&\times
 g \left ( \hat{\cal Q}^{(\mu)}\mathcal{L}_{MB}\hat{\cal P}^{(\mu)}  \right ) \nonumber  \\
&\equiv& \sum_{\xi=0}^\infty g^{\xi + 1} {\cal C}_{\xi +1}^{(\mu)}(z)  \;.
\end{eqnarray}
The lower order expansion of the creation operator reads
\begin{subequations}\label{C1C2}
\begin{eqnarray}
\hat{\mathcal{C}}^{(\mu)}_1(z)&=&{1 \over z-\hat{\cal Q}^{(\mu)}\mathcal{L}_0 \hat{\cal Q}^{(\mu)}}  \hat{\cal Q}^{(\mu)}\mathcal{L}_{MB}\hat{\cal P}^{(\mu)}   \;, \\
\hat{\mathcal{C}}_2^{(\mu)}(z)&=&{1 \over z-\hat{\cal Q}^{(\mu)}\mathcal{L}_0 \hat{\cal Q}^{(\mu)}}  \hat{\cal Q}^{(\mu)} \mathcal{L}_{MB} \hat{\cal Q}^{(\mu)}  \nonumber\\
&&\times   
 {1 \over z-\hat{\cal Q}^{(\mu)}\mathcal{L}_0 \hat{\cal Q}^{(\mu)}}  \hat{\cal Q}^{(\mu)}\mathcal{L}_{MB}\hat{\cal P}^{(\mu)}    \; .
 \end{eqnarray}
\end{subequations} 

Substituting Eq.(\ref{Cexpand}) into Eq.(\ref {App:Psi}) we have the expansion of  $\hat{\Psi}^{(\mu)}(z)$ as
\begin{equation}
\hat{\Psi}^{(\mu)}(z)={\hat{\cal P}}^{(\mu)} \mathcal{L}_0 \hat{\cal P}^{(\mu)} 
+g^2 \hat\psi_2^{(\mu)}(z) +O(g^4)  \;,
\end{equation}
where
\begin{equation}\label{psi2}
\hat{\psi}^{(\mu)}_2(z)=  \hat{\cal P}^{(\mu)}\mathcal{L}_{MB}\hat{\cal Q}^{(\mu)}  \hat{\mathcal{C}}^{(\mu)}_1(z) \;.
\end{equation}

On the other hand, the non-privileged component ($\hat{\cal Q}^{(\mu)}$-component) is obtained as a functional of the privileged component by using the creation operator as shown in Eq.(\ref {QF}).
The eigenstate $|F_{j}^{(\mu)})\!)$ is then written as
\begin{equation}\label{App:Fright}
|F_{j}^{(\mu)})\!)=\sqrt{N_{j}^{(\mu)}} \Big(\hat{\cal P}^{(\mu)}+\hat{\mathcal{C}}^{(\mu)}(Z_{j}^{(\mu)}) \Big) |u_{j}^{(\mu)})\!) \; .
\end{equation}
Since the nonequilibrium stationary state $|F_{0}^{(0)})\!)$ is a zero eigenstate of $\cal L$, it is represented by the zero eigenstate of the collision operator $|u_{0}^{(\mu)})\!)$ as
\begin{equation}\label{Fright}
|F_0^{(0)})\!)=\sqrt{N_0^{(0)}}\Big( \hat {\cal P}^{(0)}+\hat{\mathcal C}^{(0)}(+i0)\Big)|u_{0}^{(\mu)})\!) \; ,
\end{equation}
where the direction of the analytic continuation of $\hat{\cal C}^{(0)}$ is indicated by $z=+i0$, which is consistent with the fact that the approach to equilibrium is achieved in our future.
In the weak coupling case considered here, the eigenstate  $|F_0^{(0)})\!)$ is represented by the expansion of the interaction up to the second order as
\begin{equation}
\label{F00}
|F_0^{(0)})\!)=\sqrt{N_0^{(0)}} \left( 1+g\hat{\mathcal C}_{1}^{(0)}(+i0)+g^{2}\hat{\mathcal C}_{2}^{(0)}(+i0)\right)  |u_{0}^{(\mu)})\!) \;,
\end{equation}
where $\hat{\mathcal C}_{1}^{(0)}$ and $\hat{\mathcal C}_{2}^{(0)}$ are given in Eqs.(\ref{C1C2}). \par
As shown in Appendix  \ref{App:Complex}, the left eigenstate is similarly determined by the expansion of the interaction up to the second order as
\begin{equation}\label{F00Left1}
(\!(\tilde F_0^{(0)}|=\sqrt{N_0^{(0)}} (\!(\tilde v_{0}^{(0)}|\left( 1+g\hat{\mathcal D}_{1}^{(0)}(+i0)+g^{2}\hat{\mathcal D}_{2}^{(0)}(+i0)\right)   \;,
\end{equation}
where $\hat{\cal D}^{(0)}_{1}(z)$ and $\hat{\cal D}^{(0)}_{2}(z)$ are respectively the first and second order destruction-of-correlation operators which are determined by the series expansion of $g$  from  $\hat{\cal D}^{(0)}(z)$ in  Eq.(\ref {App:Destruction}).
Moreover, $(\!(\tilde v_{0}^{(0)}|$ is the left eigenstate of the collision operator shown in Eq.(\ref {App:EVPColLeft}), and the normalization constant  $N_{0}^{(0)}$ is given by Eq.(\ref {App:Nnorm}).

In order to determine $|u_{0}^{(\mu)})\!)$, we shall now solve the eigenvalue problem of the collision operator Eq.(\ref {App:EVPCol}).
Since the number of degrees of freedom of the thermal phonon system is infinitely larger than the number of molecular system $\cal N$, 
 it can be shown that the $\hat{\cal P}^{(0)}$-component of the phonon systems does not change in time and stay in their initial canonical distributions characterized by the initial temperatures $T_{L}$ and $T_{R}$.
Therefore we search for the eigenstate in the form of the tensor product of the density matrices of the molecular system and the phonon systems as
\begin{equation}\label{uj0}
|u_{j}^{(0)})\!)=|\varphi_{j}^{(0)}\>\!\>|\rho_{L}^{eq}\rho_{R}^{eq}\>\!\>  \;,
\end{equation}
where $|\rho_{L}^{eq}\rho_{R}^{eq}\>\!\>$ is a tensor product of the Liouville space vector of the left and right thermal phonon equilibrium distributions represented by Eq.(\ref{EqBath}).
In Eq.(\ref{uj0}), $|\varphi_{j}^{(0)}\>\!\>$ is an eigenstate of the reduced collision operator for the molecular system defined by
\begin{equation}
\bar{\Psi}^{(0)}(z)\equiv {\rm Tr}_{L\otimes R}\left[\hat\Psi^{(0)}(z) |\rho_{L}^{eq}\rho_{R}^{eq}\>\!\>\right]   \;,
\end{equation}
where ${\rm Tr}_{L\otimes R}$ stands for taking a partial trace of the thermal phonon systems, and therefore $\bar{\Psi}^{(0)}(z)$ is still a super-operator working on the molecular system. 

Up to the second order of the interaction, the reduced collision operator is represented by
\begin{eqnarray}\label{Psi2Expression} 
\bar{\Psi}^{(0)}_{2}&=&g^{2} \Tr_{L\otimes R} \left[\hat{\cal P}^{(0)}\mathcal{L}_{MB}\hat{\cal Q}^{(0)}  \frac{1}{i0^+-\mathcal{L}_0}  \right.
\nonumber \\
&&\times \left. \hat{\cal Q}^{(0)} \mathcal{L}_{MB}\hat{\cal P}^{(0)}|\rho_{L}^{eq}\rho_{R}^{eq}\>\!\>\right] \;.
\end{eqnarray}
There are eight diagrams for $\bar{\Psi}^{(0)}_{2}$ in our model as shown in Appendix \ref{App:DerivKinEq}.

In the present model, the matrix element of  $\bar{\Psi}^{(0)}_{2}$ is explicitly represented in terms of the transition probabilities as 
\begin{subequations}\label{RedPsi2}
\begin{eqnarray}
&&\<\!\<0,Y|\bar\Psi_2^{(0)}|0,Y\>\!\>=-i\sum_{r=L,R}\sum_{\eta >0} \Big(k_{r}^{Y-\eta,Y} 
 +k_{r}^{Y+\eta,Y} \Big)   \;,\nonumber \\
 &&\\
&& \<\!\<0,Y|\bar\Psi_2^{(0)}|0,Y+\eta\>\!\>=i k_{r}^{Y,Y+\eta}  \;,\\
 &&\<\!\<0,Y|\bar\Psi_2^{(0)}|0,Y-\eta\>\!\> = i k_{r}^{Y,Y-\eta}  \;,
\end{eqnarray}
\end{subequations}
where $k_{r}^{Y,Y\pm\eta}$  is a transition probability from $|0,Y\pm\eta\>\!\>$ ($\eta>0$) to the  $|0,Y\>\!\>$ given by Eqs.(\ref {App:tranProb1}) and (\ref {App:tranProb2}).

When we define the reduced density operator as
\begin{equation}
\label{RhoM}
f(t)  \equiv  \mathrm{Tr_{L\otimes R}}\Big[|\rho(t)\>\!\>\Big] \;,
\end{equation}
it is found that for the weakly coupled system the $f(t)$ obeys the following Pauli master equation of
\begin{eqnarray}
\label{dtfj}
\frac{d}{dt}f_{Y}^{(0)}(t)&=&-\sum_{\eta \neq 0}\sum_{r=L,R}\Big\{ k_{r}^{Y+\eta,Y}f_{Y}^{(0)}(t)  \nonumber\\
&& - k_r^{Y,Y+\eta} f_{Y+\eta}^{(0)}(t) \Big\}  \;,
\end{eqnarray}
where we have defined as
\begin{equation}\label{fY0Def}
f_{Y}^{(0)}(t)\equiv \<\!\<0,Y|f(t)\>\!\>  \;.
\end{equation}

It has been known that this type of the master equation has a unique zero eigenstate and the other eigenstates have negative imaginary values of their eigenvalues.\cite{Keizer72,Schnakenberg76,ModernThermo,HakenBook}
The zero value right eigenstate of $\bar{\Psi}_{2}^{(0)}$ is called {\it collision invariant} which is represented by
\begin{equation}
\label{ColInv}
|\varphi_0^{(0)}\>\!\> =\sum_{Y}\phi_{Y}|0,Y\>\!\>  \;.
\end{equation}
Here we take the normalization condition for $\phi_{Y}$ as
\begin{equation}
\sum_{Y}\phi_{Y}=1   \;,
\end{equation}
which gives from Eq.(\ref{uj0}) that $\mathrm{Tr}\Big[ |u_{0}^{(0)})\!) \Big] =1$.
Correspondingly the collision invariant for the left eigenstate is given by
\begin{equation}\label{v00-2}
(\!(\tilde v_{0}^{(0)}|=\sum_{Y}\sum_{\{N\}}\<\!\<0,Y|\<\!\<\{0\},\{N\}| \;,
\end{equation}
which satisfies $(\!(\tilde v_{0}^{(0)}|u_{0}^{(0)})\!)=1$.

Substituting Eq.(\ref{ColInv}) into Eqs.(\ref{uj0}) and (\ref{F00}), the nonequilibrium stationary state is obtained up to the second order of the interaction as 
\begin{eqnarray}\label{F00-2}
|F_0^{(0)})\!)&=&\sqrt{N_0^{(0)}}\left( 1+g\hat{\mathcal C}_{1}^{(0)}(+i0)+g^{2}\hat{\mathcal C}_{2}^{(0)}(+i0)\right)  \nonumber\\
&&\times \sum_{Y}\phi_{Y}|0,Y\>\!\>\prod_{r=L,R}|\rho_{r}^{eq}\>\!\>  \;.
\end{eqnarray} 
Similarly the left eigenstate for the stationary state can be obtained as
\begin{eqnarray}\label{F00Left}
(\!( \tilde F_{0}^{(0)}|&=&\sqrt{N_0^{(0)}}\sum_{Y}\<\!\<0,Y|\sum_{\{N\}}\<\!\<\{0\},\{N\}| \nonumber\\
&&\times\left( 1+g\hat{\mathcal D}_{1}^{(0)}(+i0)+g^{2}\hat{\mathcal D}_{2}^{(0)}(+i0)\right) \;,
\end{eqnarray}
which satisfies
$
 (\!(\tilde F_{0}^{(0)}|F_{0}^{(0)})\!)=1  
$
and
\begin{equation}\label{F00Norm3}
\mathrm{Tr}\Big[F_{0}^{(0)}\Big]=\sqrt{N_0^{(0)}}  \;.
\end{equation}

First we shall consider the case for ${\cal N}=2$ where the nonequilibrium stationary state is analytically obtained.
The nonequilibrium stationary population $\phi_{Y}$ ($Y=1, 2$) are obtained by
\begin{subequations}
\label{N2 Collision Invariant}
\begin{eqnarray}
\phi_{1}&=&{v_L^2 ( n_{L}(\Delta)+1) +v_R^2 ( n_{R}(\Delta)+1)   \over  v_L^2 (2 n_{L}(\Delta)+1) +v_R^2 (2 n_{R}(\Delta)+1) }  \;,  \\
\phi_{2}&=&{v_L^2  n_{L}(\Delta) +v_R^2  n_{R}(\Delta)   \over  v_L^2 (2 n_{L}(\Delta)+1) +v_R^2 (2 n_{R}(\Delta)+1) }\;,\end{eqnarray}
\end{subequations}
where $\Delta$ is given as an energy differnce between the two molecular states of $|E_{\bar 2}\>$ and $|E_{\bar 1}\>$ (See Eq.(\ref{DelDef})):
\begin{equation}
\label{DeltaDef}
\Delta\equiv\Delta_{{1},{3\over2}}={ E_{\bar 2}-E_{\bar 1}\over \hbar }= {1\over\hbar}\sqrt{(\varepsilon_{L}-\varepsilon_{R})^{2}+4 J^{2}} \;.
\end{equation}

\begin{figure}
\begin{center}
\includegraphics[width=8cm]{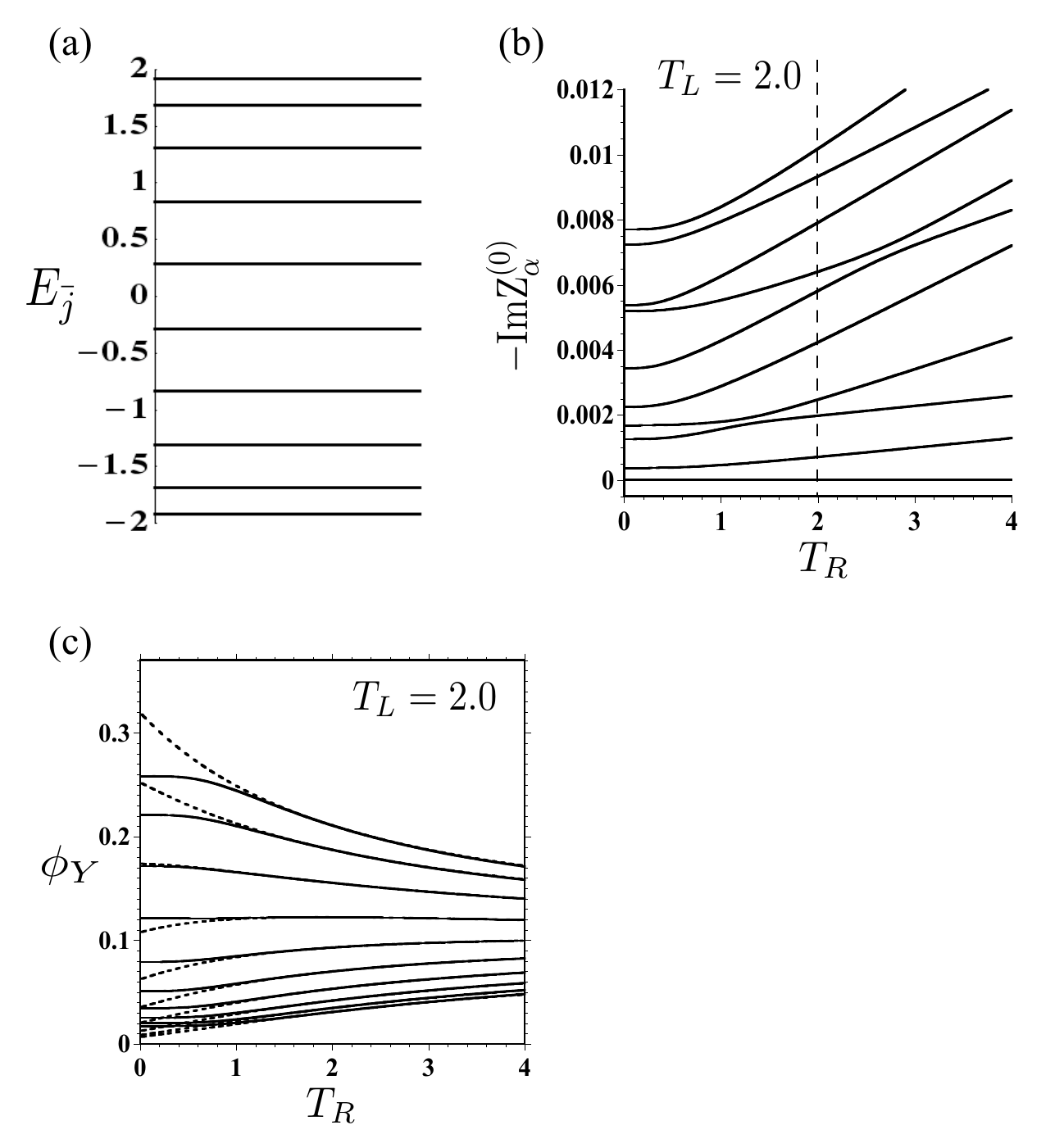}
\caption{(a) The molecular level structure for the chain with ${\cal N}=10$, where the parameters are $\varepsilon_{L}=\varepsilon_{R}=0, J=1.0, v_{R}=v_{L}=v=1.0$, and $g=0.1$.
(b)  The spectrum of the collision operator $Z_{\alpha}^{(0)}$ as a function of $T_{R}$ changing from 0 to 4.0 with $T_{L}=2.0$ fixed, where the vertical axis denotes $-\rm{Im}Z_{\alpha}^{(0)}$.
(c) The state population of the nonequilibrium stationary state $\phi_{Y}$ as a function of $T_{R}$.
The dotted lines are the population for the average temperature of $T_{M}=(T_{L}+T_{R})/2$.
 }
\label{Fig:Kspec} 
\end{center}
\end{figure}

For a longer molecular chain, it is difficult to analytically obtain the nonequilibrium stationary state.
Instead we have numerically solved the eigenvalue problem of the collision operator and obtained the collision invariant.
As an example, we show in Fig.\ref{Fig:Kspec} the results for a molecule with ${\cal N}=10$:
the molecular level structure, the eigenstates of the molecular Hamiltonian (a), the spectrum of the collision operator (b), and the population of the collision invariant (c).
In Fig.\ref {Fig:Kspec}(b) and (c) we have fixed at $T_{L}=2.0$ and changed $T_{R}$ and have taken  $ v_{L}=v_{R}=1$ and $g=0.1$.
Since there are ten basis states belonging to $P^{(0)}$ subspace ( $|0,Y\>\!\>$ with  $Y={1}$ to ${10}$ ), we have ten eigenstates of the collision operator.
For any $T_{R}$, there always exists an eigenstate with $Z_{0}^{(0)}=0$, {\it i.e.} a collision invariant, while other eigenstates are decaying states with $\rm{Im}Z_{\alpha(\neq 0)}^{(0)}<0$, as mentioned above.

In Fig.\ref{Fig:Kspec}(c), we have also shown by the dotted lines the canonical distribution for the average temperature $T_{M}\equiv (T_{L}+T_{R})/2$.
At the low temperature, the nonequilibrium state population is different from the canonical distribution for the average temperature, while both agree well at the higher temperature.
Furthermore, at the low temperature case where the discrete molecular level structure plays a key role, the nonequilibrium state population cannot be represented by the canonical distribution for any unique temperature.
This suggests that due to the quantum effect the local temperature of the molecule cannot be identified at the low temperature case, in contrast to the classical system.\cite{Rego98}

\section{Energy Flow}\label{Sec:EFlow}

In this section, we consider the energy flow in the nonequilibrium stationary state in terms of the zero eigenstate of  $\cal L$ obtained in Section \ref {Sec:Eigen}.
The energy flow through the molecule is evaluated by considering the energy change of the molecule.
The energy change of the molecule in the nonequilibrium stationary state is obtained by
\begin{equation}
\label{Wdef}
w(t)\equiv \frac{d}{dt}\<H_M\>_t= \frac{d}{dt}{\rm Tr}\left[H_{M}\rho(t)\right]  \;.
\end{equation}

Using Eq.(\ref {Liouville}), $w(t)$ is represented by
\begin{subequations}\label{WtRW}
\begin{eqnarray}
w(t)
&=&-i\<\!\<H_M|{\rm Tr}_{L\otimes R}\big[{\cal L}\rho(t)\big]  \>\!\> \\
&=&-i g \<\!\< H_M|{\rm Tr}_{L\otimes R}\big[{\cal L}_{MB} \rho(t)\big] \>\!\>  \;,
\end{eqnarray}
\end{subequations}
where we have used the fact that  $\left( {\cal L}_{M} H_{M}\right)^{\dagger}=0$ and ${\rm Tr}_{L\otimes R}[{\cal L}_{B}\rho(t)]=0$ in the second equality.
Taking into account ${\cal L}_{MB}={\cal L}_{MB}^{L}+{\cal L}_{MB}^{R}$ corresponding to $H_{MB}=H_{MB}^{L}+H_{MB}^{R}$ in Eq.(\ref{HMB}), we can divide $w(t)$ into two contributions due to the interactions with the left and the right thermal baths:
\begin{eqnarray}
\label{WLWR}
w(t)=w^{L}(t)+w^R(t) \;,
\end{eqnarray}
where
\begin{equation}
\label{WBdef}
w^{r}(t)\equiv-i g \<\!\<H_M|{\rm Tr}_{L\otimes R}\left[{\cal L}_{MB}^{r} \rho(t)\right] \>\!\> \quad (r=L,R)   \;.
\end{equation}
The sign of $w^{r}(t)$ is positive when energy flow comes into the molecule from the thermal bath $r=L$ or $R$. 

Since  ${\cal L}F_{0}^{(0)}=0$ in the nonequilibrium stationary state, we find that $w (\infty)=-i\<\!\<H_M|{\rm Tr}_{L\otimes R}\big[{\cal L}F_{0}^{(0)}\big]  \>\!\>=0$, resulting in $w^{L}(\infty )=-w^{R}(\infty )$ which guarantees that  influx and outflow of molecular energy are balanced in the stationary state.
We then define the energy flow from the left to the right thermal bathes going through the molecule as 
\begin{equation}\label{Energy Flow Def}
{\cal U}(t)\equiv {1\over 2}\left( w^{L}(t)-w^{R}(t)\right)  \;.
\end{equation}

Since in the long time Eq.(\ref{F00Norm3}) indicates that
\begin{equation}\label{rhoLim}
|\rho(t) )\!) \xrightarrow{t\rightarrow\infty} {1\over \sqrt{N_0^{(0)}}}|F_{0})\!) \;,
\end{equation}
it is found from Eq.(\ref{F00-2}) that ${\cal U}(\infty)$  can be represented up to the second order of ${\cal L}_{MB}$ as 
\begin{equation}
\label{Wst2}
{\cal U}(\infty)= -i g^{2}\<\!\< H_M|\mathrm{Tr}_{L\otimes R} \left[{\cal L}_{MB}^{L}\hat{\cal C}_{1}^{(0)}(+i0)\varphi_0^{(0)} \rho_{L}^{eq}\rho_{R}^{eq}\right] \>\!\>  \;. 
\end{equation}
It should be noted that the first order correlation created from the collision invariant contributes to  the energy flow in the nonequilibrium stationary state.\cite{PrigogineText,ResiboisBook,Resibois67}
Furthermore, by comparison with Eq.(\ref {psi2}), it is found that the energy flow is related to the collision operator which is expressed by the transition probability given in Eq.(\ref{RedPsi2}).

Inserting Eqs.(\ref{ColInv}) and (\ref {C1C2}a) into (\ref{Wst2}) and using the matrix representation of ${\cal L}_{MB}^{L}$ given in Eqs.(\ref {App:LMBcomp}), we can explicitly represent ${\cal U}^{st}\equiv{\cal U}(\infty)$ as a sum of any pair of the states of $|0,Y\>\!\>$ and $|0,Y+\eta\>\!\>$:
\begin{subequations}\label{WstExp}
\begin{eqnarray}
&&{\cal U}^{st}\equiv  \sum_{\eta>0}\sum_{Y} {\cal U}_{Y+\eta,Y} \nonumber\\
&&=\frac{4\pi g^2 v_L^2}{\hbar} \sum_{\eta>0}\sum_{Y}{\cal D}^{ph}(\Delta_{\eta,Y+{\eta\over2}})\big|c_{L,Y+\eta}^{*}c_{L,Y}\big|^2  \Delta_{\eta,Y+{\eta\over2}} \nonumber\\
&&\times\bigg\{ \phi_{Y}n_L(\Delta_{\eta,Y+{\eta\over2}})-\phi_{Y+\eta} \big( n_L(\Delta_{\eta,Y+{\eta\over2}})+1 \big) \bigg\} \nonumber\\
&&\\
&&=2 \sum_{\eta>0}\sum_{Y}\hbar  \Delta_{\eta,Y+{\eta\over2}}\Big\{  k_L^{Y+\eta,Y} \phi_{Y}-k_L^{Y,Y+\eta} \phi_{Y+\eta} \Big\} \;, \nonumber \\ 
\end{eqnarray}
\end{subequations}
where $2\textstyle\sum_{\eta>0}$ in Eq.(\ref{WstExp}b) can be replaced by $\textstyle\sum_{\eta\neq0}$.
Here we would like to make a comment that the same formula can be derived by evaluating energy change of the thermal bath as shown in Appendix \ref{App:BathChange}.

We note that there is a striking correspondence between Eq.(\ref{WstExp}b) and Eq.(\ref{dtfj}): The right hand side of the two equations involve a common factor of $(  k_L^{Y+\eta,Y} \phi_{Y}-k_L^{Y,Y+\eta} \phi_{Y+\eta} )$ which is the transition rate per time between the states of $|0,Y\>\!\>$ and $|0,Y+\eta\>\!\>$ due to the coupling with the left thermal bath.
The energy flow coming in from the left thermal bath is obtained by multiplying with it the energy difference between the states of $|0,Y\>\!\>$ and $|0,Y+\eta\>\!\>$ states, $\hbar\Delta_{\eta,Y+\eta/2}$.
This correspondence is originated in the fact that the collision operator $\hat\psi_{2}^{(0)}(z)$ can be represented by the first order creation operator $\hat{\cal C}_{1}^{(0)}(z)$ as shown in Eq.(\ref{psi2}).
Therefore in the weak coupling case the energy flow can be derived by using the solution of the eigenvalue problem of the collision operator: Simply multiplying the energy difference of the states and the transition rate between them.

This correspondence naturally leads it to a Landauer formula which has been widely used to interpret the carrier flow of the mesoscopic system in nonequilibrium situation.\cite{Landauer57,Rego98,Schwab00,Wharam88}
Landauer formula reads
\begin{equation}\label{wst}
{\cal U}^{st}=\int_{0}^{\infty} \hbar \omega{\cal T}(\omega ) \big(n_{L}(\omega)-n_{R}(\omega)\big)   d\omega\;,
\end{equation}
where ${\cal T}(\omega )$ is  a transmission function determined by using Fermi's golden rule.
In the simplest approximation, ${\cal T}(\omega)=1$ is assumed, which results in a quantization of thermal conductance.\cite{Rego98}
In more elaborate works, they have estimated the $\omega$ dependence of ${\cal T}(\omega)$ reflecting a resonance effect due to the discretized molecular level structure.\cite{Segal01,Segal03}

In the present model, the equation (\ref {WstExp}) can be cast into the form of Eq.(\ref {wst}) in terms of the transmission function defined by
\begin{eqnarray}\label{TransmissionF}
&&{\cal T}(\omega)= 2 \sum_{\eta>0}\sum_{Y} \delta (\omega-\Delta_{\eta,Y+{\eta\over 2}})  \nonumber\\
& & \times{ (k_L^{Y+\eta,Y}-k_R^{Y+\eta,Y} )\phi_{Y}- (k_L^{Y,Y+\eta}-k_R^{Y,Y+\eta} )\phi_{Y+\eta}  \over  n_L(\Delta_{\eta,Y+\eta})-  n_R(\Delta_{\eta,Y+\eta})  }  \;. \nonumber\\
&&
\end{eqnarray}
The transmission function has a strong resonance characteristic for the molecular level structure, which is reflected in the delta function of Eq.(\ref{TransmissionF}): $\omega=(E_{Y+\eta}-E_{Y})/\hbar$.


While the energy flow due to the first order correlation can be represented by the Landauer formula as shown above, 
we shall show in the next section a physical quantity due to the higher order correlation, such as an induced polarization, cannot be represented by the Landauer formula.
It is worthwhile to note that the Landauer formula is derived in our approach through the resonance effects between the molecular chain and the thermal baths, i.e., the dissipation occurs at the edges of the system contacting to the baths.

Before going to the next section, we shall show some examples of the energy flow in the nonequilibrium stationary state of the molecule. 
The energy flow is analytically obtained for a molecule with the length of ${\cal N}=2$ by using Eq.(\ref {N2 Collision Invariant}).
In this case, we have
\begin{eqnarray}\label{T2site}
{\cal T}(\omega)
&=&\delta(\omega-\Delta){2\pi g^{2} \over \hbar^2} {4 J^2 \over (\varepsilon_L-\varepsilon_R)^2 + 4 J^2} \; {\cal D}_{ph}(\Delta) \nonumber\\
&\times&{v_L^2 v_R^2\over (2 n_L(\Delta)+1)v_L^2+(2 n_R(\Delta)+1)v_R^2 } \;,
\end{eqnarray}
yielding 
\begin{eqnarray}\label{N2 energy flow}
{\cal U}^{st}&=&{2\pi  g^2\over \hbar} { J^2 \over (\varepsilon_L-\varepsilon_R)^2 + 4 J^2} \; {\cal D}_{ph}(\Delta)\nonumber\\
&\times& {2 v_L^2 v_R^2 (n_L(\Delta)-n_R(\Delta))\over (2 n_L(\Delta)+1)v_L^2+(2 n_R(\Delta)+1)v_R^2 }
\end{eqnarray}
where $\Delta$ is given by Eq.(\ref{DeltaDef}).

\begin{figure}
\includegraphics[width=8cm]{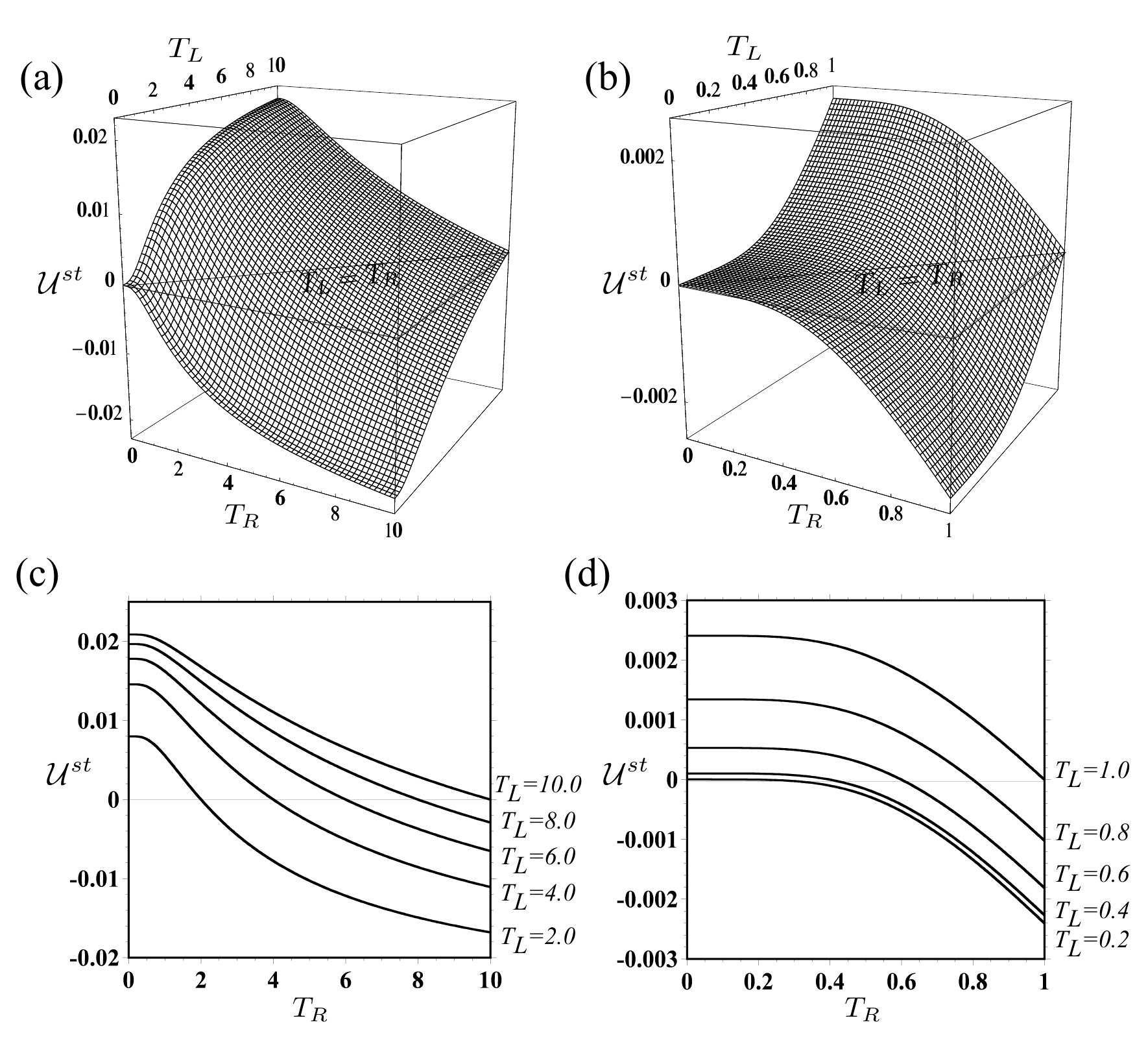}
\caption{The energy flow ${\cal U}^{st}$ of a molecule with the length of ${\cal N}=10$ in the $T_{L}$-$T_{R}$ plane.
The parameters for the calculation are the same as used in Fig.\ref{Fig:Kspec}.
The temperature range is taken from 0 to 10 in (a), and from 0 to 1 in (b).
The energy flow as a function of $T_{R}$ for several fixed values of $T_{L}$ are shown in (c) and (d).}
\label{Fig:EFlow} 
\end{figure}

For a longer molecule we have numerically calculated the energy flow.
As an example, we have shown in Fig.\ref{Fig:EFlow} the energy flow through the same molecule as studied in Section \ref {Sec:Eigen}. 
We show the energy flow ${\cal U}^{st}$ in the bird's-eye view in the $T_{L}$-$T_{R}$ plane in Fig.\ref{Fig:EFlow}(a) and (b).
The wide temperature region of ${\cal U}^{st}$ is shown in (a), and the low temperature region in an expanded scale in (b).
We also show the energy flow ${\cal U}^{st}$ as a function of $T_{R}$ for various values of $T_{L}$ in (c) and (d).
It should be noted that while the energy flow is linearly proportional to the temperature difference near thermal equilibrium $T_{L}\simeq T_{R}$, this linear relation breaks down in far-from equilibrium.

\begin{figure}
\begin{center}
\includegraphics[width=8cm]{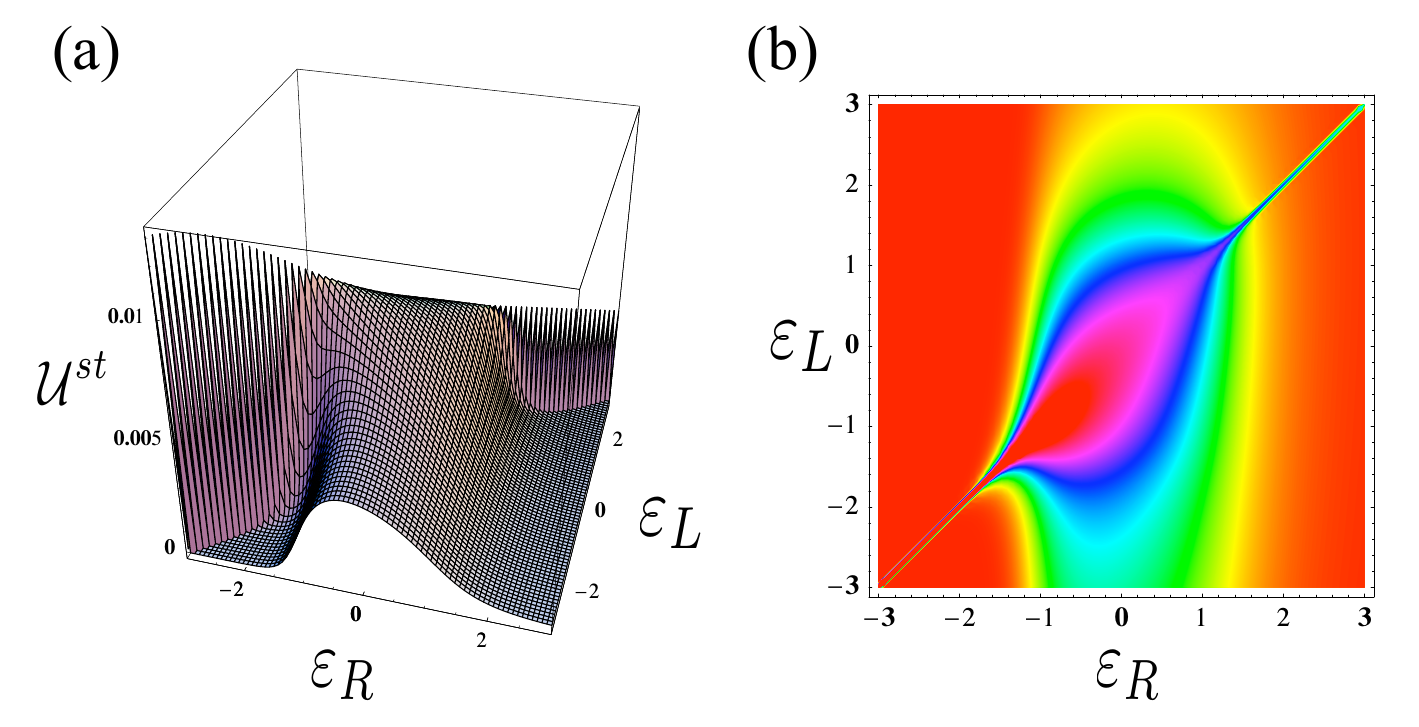}
\caption{The dependence of the energy flow ${\cal U}^{st}$ on $\varepsilon_{L}$ and $\varepsilon_{R}$ with $T_{L}=2.0$ and $T_{R}=0.01$ fixed, where the other parameters are the same as used in Fig.\ref{Fig:EFlow}.
The bird's eye view and the contour plot are shown in (a) and (b), respectively.
}
\label{Fig:ELER} 
\end{center}
\end{figure}

The energy flow ${\cal U}^{st}$ also depends on $\varepsilon_{L}$ and $\varepsilon_{R}$.
We show in  Fig.\ref{Fig:ELER} the energy flow ${\cal U}^{st}$ as a function of $\varepsilon_{L}$ and $\varepsilon_{R}$ for  $T_{L}=2.0$ and $T_{R}=0.01$, where the other parameters are the same as in Fig.\ref{Fig:EFlow}.
The bird's eye view and the contour plot are shown in (a) and (b), respectively.
The ${\cal U}^{st}$ is maximized at $\varepsilon_{L}-\varepsilon_{R}=0$ as a function of   $\varepsilon_{L}-\varepsilon_{R}$,  because a transfer between the end states $|L\>$ and $|R\>$ through the molecule occurs most effectively.
For different values of  $\varepsilon_{L}$ and $\varepsilon_{R}$, the energy transfer is allowed to occur due to the energetically spread  molecular states as shown in Fig.\ref{Fig:Scheme}.
As temperature increases, the energy flow increases.

\section{Induced Polarization, Particle current}\label{Sec:Pflow}

In spite of the fact that the thermal force stems from complicated many-body dissipative effect, when the polarizable molecule is subject to an external thermal force under nonequilibrium condition, a polarization is induced, just like a simple mechanical force with an external electric field.
We shall show that an induced polarization within the molecular states is attributed to the second order correlation so that the quantity cannot be reduced in a Landauer formula unlike the energy flow.

Polarization operator is represented by
\begin{equation}
\label{Polarization}
e \hat x=\sum_{m=1}^{\cal N} e x_{m} |m\>\<m|  \;,
\end{equation}
when  $x_{m}=m d$ with a lattice constant $d$ and the electric charge $e$.
In the present work, we take $e=1$.
The conjugate current to the polarization is defined as a time derivative of the polarization:
\begin{equation}
{\cal I}(t)\equiv \frac{d}{dt}\<\hat{x}\>_t= {d\over dt}{\rm Tr}[ \hat x \rho(t) ]  \;.
\end{equation}
Using Eq.(\ref {Liouville}), ${\cal I}(t)$ is then represented as
\begin{eqnarray}
\label{Current}
{\cal I}(t)\equiv \frac{d}{dt}\<\hat{x}\>_t&=&-i \<\!\< \hat x|{\rm Tr}_{L\otimes R}[\mathcal L \rho(t) ] \>\!\>  \nonumber\\
&=&-i\<\!\< \hat x| \mathcal L_{M} |f(t)  \>\!\>  \;,
\end{eqnarray}
where we have used the fact that $\<\!\< \hat x|{\rm Tr}_{L\otimes R} [{\cal L}_{B}+g {\cal L}_{MB}]=0$, and $f(t)$ is a reduced density operator of the particle given by Eq.(\ref {RhoM}).
When we define the current operator as
\begin{equation}
\hat{\cal I}\equiv i[H_{M},\hat x] \;,
\end{equation}
then the current ${\cal I}(t)$ is given by ${\cal I}(t)=\mathrm{Tr}[\hat{\cal I} f(t)]$.
Note that the current operator $\hat{\cal I}$ may be represented by using the site basis $\{ |m\>\}$  as
\begin{eqnarray}
\label{CurOp}
&&\hat{\cal I} =\frac{i}{\hbar}\sum_{m=1}^{N} x_m\big[ H_M,|m\>\<m|\big] \nonumber \\
&&= - \frac{i  d}{\hbar} \sum_{m=1}^{N-1} J_{m,m+1} \left(|m+1\>\<m|-|m\>\<m+1|\right)   \;,\nonumber\\
\end{eqnarray}
which agrees with the ordinary definition of the particle current in a one-dimensional  discrete lattice.\cite{MahanBook}

Considering that $\rho(t)$ in the long time limit is given by $F_{0}^{(0)}$ Eq.(\ref{rhoLim}), we then have
\begin{equation}\label{TotalI}
{\cal I}^{st}\equiv -i \<\!\< \hat x|{\rm Tr}_{L\otimes R} [\mathcal LF_{0}^{(0)}] \>\!\> =0  \;,
\end{equation}
i.e.,  the total particle current in the nonequilibrium stationary state vanishes because  ${\cal L}F_{0}^{(0)}=0$. 

Substituting the spectral representation of ${\cal L}_{M}$ given by Eq.(\ref {LMEV}) into Eq.(\ref {Current}),  we have
\begin{subequations}\label{CurrentComp}
\begin{eqnarray}
{\cal I}(t)&=&-i\sum_{\eta(\neq 0),Y}\<\!\<\hat x|\eta,Y\>\!\>\Delta_{\eta,Y}\<\!\<\eta,Y|f(t)\>\!\> \nonumber\\
&&\\
&=&-i\sum_Y\sum_{\eta> 0}\<\!\<\hat x|\eta,Y\>\!\>\Delta_{\eta,Y} \nonumber\\
&&\quad\times \big(\<\!\<\eta,Y|-\<\!\<-\eta,Y| \big)  |f(t)\>\!\>\\
&\equiv&\sum_Y\sum_{\eta> 0}  {\cal I}_ {\eta,Y}(t)  \;,
\end{eqnarray}
\end{subequations}
where we have defined a particle current component ${\cal I}_{\eta,Y}(t)$.
The current component ${\cal I}_{\eta,Y}(t)$ is a product of the three factors: i) Transition dipole moment between $|Y+\eta\>$ and $|Y-\eta\>$ molecular states,  $\<\!\<\hat x|\eta,Y\>\!\> =\<Y-\eta/2|\hat x|Y+\eta/2\>$, ii) Transition frequency between these two states, $\Delta_{\eta,Y}$, and iii) Off diagonal matrix element of the reduced density matrix,  $(\<\!\<\eta,Y|-\<\!\<-\eta,Y|)|f(t)\>\!\> $.
The product of the factors i) and ii) represents that the particle current is attributed to the transition of the particle between $|Y-\eta/2\>$ and $|Y+\eta/2\>$ molecular states.
Furthermore, since the factor iii) is written as
\begin{eqnarray}\label{ftoffstate}
&&(\<\!\<\eta,Y|-\<\!\<-\eta,Y|\Big)|f(t)\>\!\>\nonumber\\
&&=\big\langle Y+{\eta\over2}\big|f(t)\big|Y-{\eta\over2}\big\rangle- \big\langle Y-{\eta\over2}|f(t)\big|Y+{\eta\over2}\big\rangle \;, \nonumber\\
\end{eqnarray}
 the reduced density matrix $f(t)$ should be non-Hermitian in order to obtain a non-vanishing ${\cal I}_{\eta,Y}(t)$.
This is quite contrast to a thermal equilibrium at $\beta=1/T$ where the density matrix is given by a Hermitian matrix $\rho_{eq}=\exp[- \beta H]$ which  leads to a vanishing particle current component, while ${\cal I}_{\eta,Y}^{st}$ take finite values in the nonequilibrium stationary state.

Substituting $F_{0}^{(0)}$ into  Eqs.(\ref{CurrentComp}), the current component is expressed by 
\begin{eqnarray}\label{Jrep1}
{\cal I}_{\eta,Y}^{st}&=&-i\<\!\<\hat x|\eta,Y\>\!\>\Delta_{\eta,Y}\big(\<\!\<\eta,Y|-\<\!\<-\eta,Y| \big) \nonumber\\
&\times&  g^{2}\big|{\rm Tr}_{L\otimes R}\big[ \hat{\cal C}_{2}^{(0)}(+i0)\varphi_{0}^{(0)}\rho_{ph}^{eq}\big] \>\!\> 
  \textrm{,  for  }\eta>0 \;.
\end{eqnarray}
It should be noted that the contribution to the current component ${\cal I}_{\eta,Y}^{st}$ is attributed to the second order creation operator.

Inserting Eqs.(\ref{ColInv}) and (\ref{C1C2}) into Eq.(\ref{Jrep1}), the explicit expression of ${\cal I}_{\eta,Y}^{st}$ is obtained as
\begin{widetext}
\begin{eqnarray}\label{Icomp1}
&&{\cal I}_{\eta,Y}^{st}=\frac{2\pi }{\hbar^2}\<\!\<\hat x|\eta,Y\>\!\>  \sum_{r=L,R} g^2 v_r^2   c_{r,Y+{\eta\over 2}}^{*}c_{r,Y-{\eta\over2}}  \nonumber\\
&&
 \times\sum_{\xi>0}\bigg\{ {\cal D}_{ph}(\Delta_{\xi,Y-\eta-\xi}) |c_{r,Y-{\eta\over2}-\xi}|^2 \big\{\phi_{Y-{\eta\over2}}\big(n_r(\Delta_{\xi,Y-\eta-\xi})+1\big)-\phi_{Y-{\eta\over2}-\xi}n_r(\Delta_{\xi,Y-\eta-\xi})\big\}   \nonumber\\
&& \qquad \quad+ {\cal D}_{ph}(\Delta_{\xi,Y-\eta+\xi}) |c_{r,Y-{\eta\over2}+\xi}|^2 \big\{\phi_{Y-{\eta\over2}}n_r(\Delta_{\xi,Y-\eta+\xi})-\phi_{Y-{\eta\over2}+\xi}\big(n_r(\Delta_{\xi,Y-\eta+\xi}) +1\big)\big\} \nonumber\\
&&
\qquad \quad+ {\cal D}_{ph}(\Delta_{\xi,Y+\eta-\xi}) |c_{r,Y+{\eta\over2}-\xi}|^2 \big\{\phi_{Y+{\eta\over2}}\big(n_r(\Delta_{\xi,Y+\eta-\xi})+1\big)-\phi_{Y+{\eta\over2}-\xi}n_r(\Delta_{\xi,Y+\eta-\xi})\big\}   \nonumber\\
&& \qquad \quad+ {\cal D}_{ph}(\Delta_{\xi,Y+\eta+\xi}) |c_{r,Y+{\eta\over2}+\xi}|^2 \big\{\phi_{Y+{\eta\over2}}n_r(\Delta_{\xi,Y+\eta+\xi})-\phi_{Y+{\eta\over2}+\xi}\big(n_r(\Delta_{\xi,Y+\eta+\xi}) +1\big)\big\} \;, 
\end{eqnarray}
or by using the transition probabilities of $k_{r}^{Y,Y+\eta}$ given by Eqs.(\ref {App:tranProb1}),  ${\cal I}_{\eta,Y}^{st}$ can be written as
\begin{eqnarray}\label{Icomp2}
{\cal I}_{\eta,Y}^{st}= \big\langle Y-{\eta\over2}\big|\hat x\big|Y+{\eta\over2}\big\rangle \sum_{r=L,R} \sum_{\xi\neq 0}&&\bigg\{{ c_{r,Y+{\eta\over2}}\over c_{r,Y-{\eta\over2}}} \big( k_r^{Y+{\eta\over2}+\xi,Y-{\eta\over2}}\phi_{Y-{\eta\over2}} - k_r^{Y-{\eta\over2},Y+{\eta\over2}+\xi}\phi_{Y+{\eta\over2}+\xi} \big)  \nonumber\\
&&+ { c_{r,Y-{\eta\over2}}\over c_{r,Y+{\eta\over2}}} \big( k_r^{Y+{\eta\over2}+\xi,Y+{\eta\over2}}\phi_{Y+{\eta\over2}} - k_r^{Y+{\eta\over2},Y+{\eta\over2}+\xi}\phi_{Y+{\eta\over2}+\xi} \big)   \bigg\} \;. 
\end{eqnarray}
\end{widetext}
As seen in Eq.(\ref{Icomp2}), in the calculation of the current component ${\cal I}_{\eta,Y}^{st}$ due to the pair of the molecular states between $|Y+\eta/2\>$ and $|Y-\eta/2\>$ we have to consider the transitions between the other molecular state $|Y+\eta'\>$ and $|Y\>$ expressed by $k_r^{Y+\eta',Y}$ or $k_r^{Y,Y+\eta'}$, and between  $|Y+\eta'\>$ and $|Y+\eta\>$ expressed by $k_r^{Y+\eta',Y+\eta}$ or  $k_r^{Y+\eta,Y+\eta'}$.
As a result, the particle current cannot be cast into the Landauer formula contrary to the energy flow,
where there is no other molecular states involved in the calculation of the energy flow ${\cal U}_{Y+\eta,Y}^{st}$.

\begin{figure}
\begin{center}
\includegraphics[width=8cm]{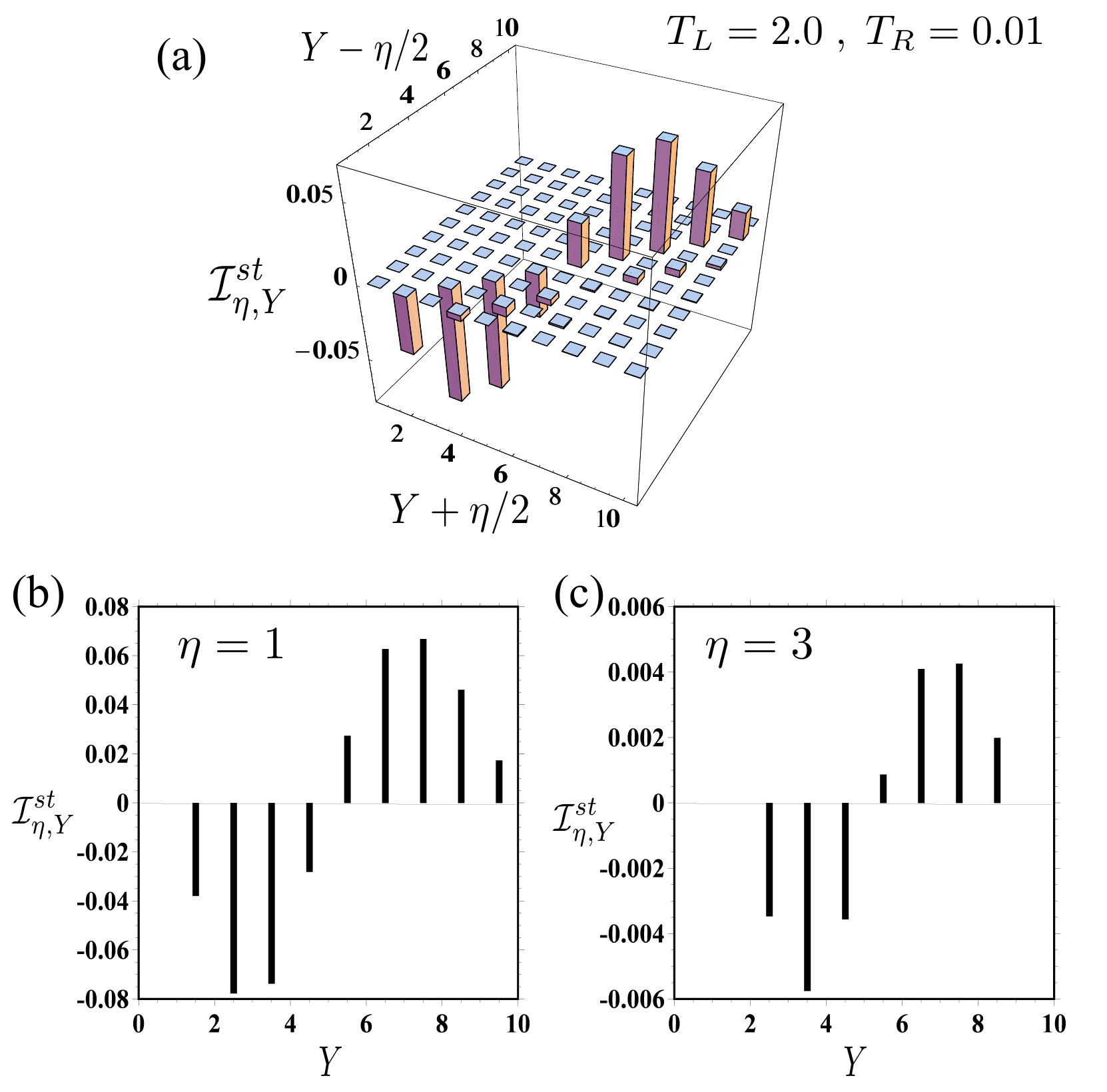}
\caption{The particle current components ${\cal I}_{\eta(>0),Y}^{st}$ for ${\cal N}=10$, $v_{L}=v_{R}=1.0$, $g=0.1$, $T_{L}=2.0$, and $T_{R}=0.01$.
In (a), all the components of ${\cal I}_{\eta,Y}^{st}$ are drawn, and in (b) and (c) the components of $\eta=1$ and $\eta=3$ are shown, respectively. 
 }
\label{Fig:Jmat1} 
\end{center}
\end{figure}
\begin{figure}
\begin{center}
\includegraphics[width=8cm]{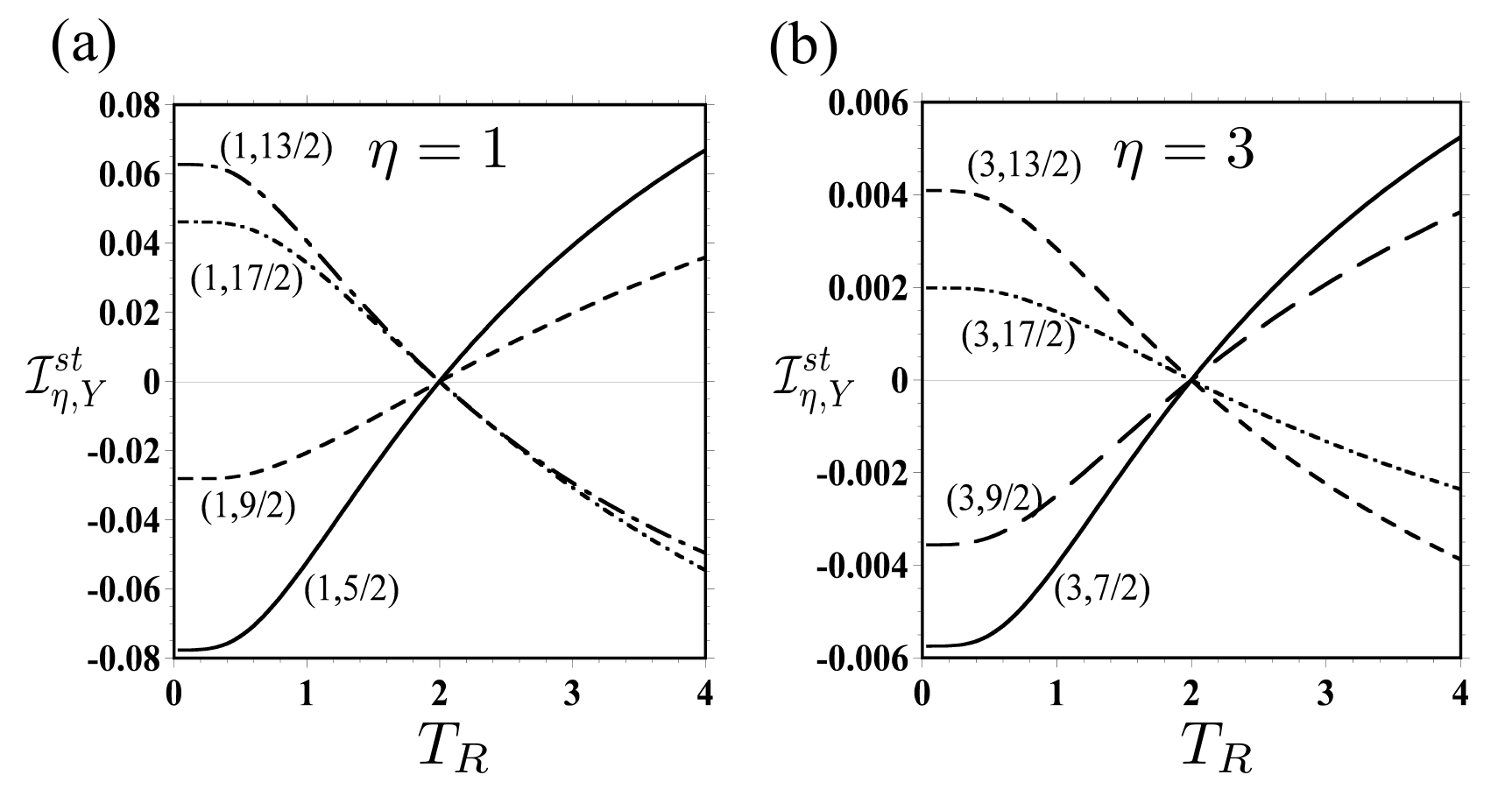}
\caption{The temperature dependence of some current components shown in Fig.\ref{Fig:Jmat1}.
The current components as a function of $T_{R}$ for a fixed value of $T_{L}=2.0$ are shown.
The ${\cal I}_{\eta,Y}$ for $(\eta,Y)=(1,5/2), (1,9/2), (1,13/2), (1,17/2)$ and $(\eta,Y)=(3,7/2), (3,9/2), (3,13/2), (3,17/2)$ are shown in (a) and (b), respectively.
 }
\label{Fig:IcomTR} 
\end{center}
\end{figure}

As an example of the particle current we show in Fig.\ref{Fig:Jmat1} the particle current components ${\cal I}_{\eta,Y}^{st}$ of the same  molecule which has been investigated in Section \ref {Sec:EFlow}, where we have taken ${\cal N}=10$, $v_{L}=v_{R}=1.0$, $g=0.1$, $T_{L}=2.0$, and $T_{R}=0.01$.
In (a), all the components of ${\cal I}_{\eta,Y}^{st}$ are drawn, and in (b) and (c) the components for fixed values of $\eta=1$ and $\eta=3$ are shown, respectively, where the horizontal axis is $Y$.
When $\varepsilon_{L}=\varepsilon_{R}$, ${\cal I}_{\eta,Y}^{st}=0$ for an even integer of $\eta$, because $\<\!\<\eta,Y|\hat x\>\!\>=0$ due to the symmetry.
It is found that ${\cal I}_{\eta,Y}^{st}$ takes a large value when $\eta=1$ which suggests that the particle current is large for a pair of  adjacent molecular states in energy, and  the ${\cal I}_{\eta,Y}^{st}$ becomes small as $\eta$ increases.
It is also found that ${\cal I}_{\eta,Y}^{st}$ is positive for a large $Y$ while it is negative for a small $Y$: The induced polarization due to the molecular states with a higher energy is directed from high temperature side to the low temperature side.

We also show the temperature dependence of ${\cal I}_{\eta,Y}^{st}$ of Fig.\ref{Fig:IcomTR} as a function of  $T_{R}$ with a fixed value of $T_{L}=2.0$.
In (a) are shown ${\cal I}_{\eta,Y}^{st}$ for $(\eta,Y)=(1,5/2), (1,9/2), (1,13/2), (1,17/2)$, and in (b) for   $(\eta,Y)=(3,7/2), (3,9/2), (3,13/2), (3,17/2)$.
The particle current vanishes at $T_{R}=T_{L}$, and they linearly depends on the temperature difference around $T_{R}=T_{L}$, while they change nonlinearly in the low temperature region reflecting the discrete molecular level structures.
This behavior corresponds to that of the energy flow shown in Fig.\ref {Fig:EFlow}, because both the energy flow and the particle current are born out from the unique collision invariant by the actions of $\hat{\cal C}_{1}^{(0)}$ and  $\hat{\cal C}_{2}^{(0)}$ on  $|u_{0}^{(0)})\!)$.

\begin{figure}
\begin{center}
\includegraphics[width=8cm]{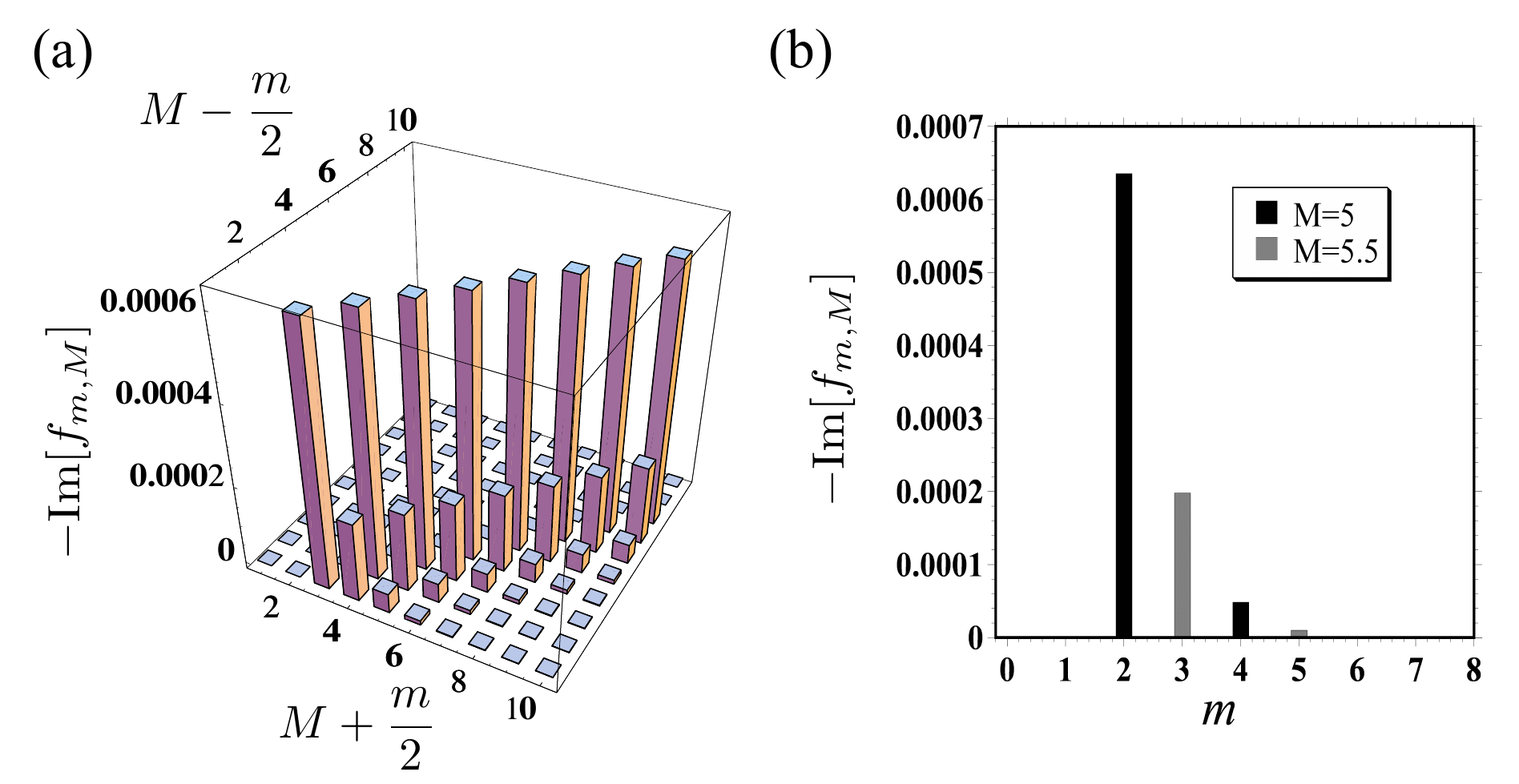}
\caption{Space correlation in the molecule. All the parameters are the same as in Fig.\ref{Fig:Jmat1}; (a) All the components of $f_{m,M}$, and (b) the dependence on the relative distance $m$ for $M=5$ and $M=5.5$. }
\label{Fig:SpaceCor} 
\end{center}
\end{figure}

So far we have investigated the particle current components attributed to a quantum correlation between a pair of molecular states.
With use of the representation in terms of the site basis given in Eq.(\ref {Barj}), we may reveal the quantum correlation in space which is generated in the nonequilibrium stationary state.
Similarly to Eq.(\ref{ftoffstate}), we shall investigate the non-symmetricity of the off-diagonal elements of $f(t)$ in terms of the site basis as
\begin{eqnarray}\label{ftm1m2}
f_{m,M}\equiv\big(\<\!\<m,M|-\<\!\<-m,M|\big) |f(t)\>\!\> \;,
\end{eqnarray}
where $|m,M\>\!\>$ is the Wigner basis in terms of the site basis defined similarly to Eq.(\ref {WignerPtcl})  by
\begin{eqnarray}
&&|\pm m,M\>\!\>\equiv |M\pm{m\over 2};M\mp{m\over 2}\>\equiv |M\pm{m\over 2}\>\<M\mp{m\over 2}|  \;,  \nonumber\\
 &&\hspace{3cm} \textrm{(double sign in same order)} 
\end{eqnarray}
with use of site basis $|M\pm{m/ 2}\>$.
We show in Fig.\ref {Fig:SpaceCor} the imaginary part of $f_{m,M}$ for the same case of Fig.\ref{Fig:Jmat1}: In (a) we show the all the components of $f_{m,M}$ and in (b) the dependence on the relative distance $m$ for $M=5$ and  $M=5.5$.
As seen from the figures, the values of $f_{m,M}$ does not depend on the central position of $M$ but on the relative distance of $m$, suggesting that the quantum correlation in space decreases, as $m$ increases.
We also found that $f_{m,M}=0$ at  $T_{L}=T_{R}$: The quantum correlation in space comes to appear only at nonequilibrium situation.

\section{Application to one-dimensional DNA chain}\label{Sec:Discussion}

\begin{figure}[tp]
\begin{center}
\includegraphics[width=8cm]{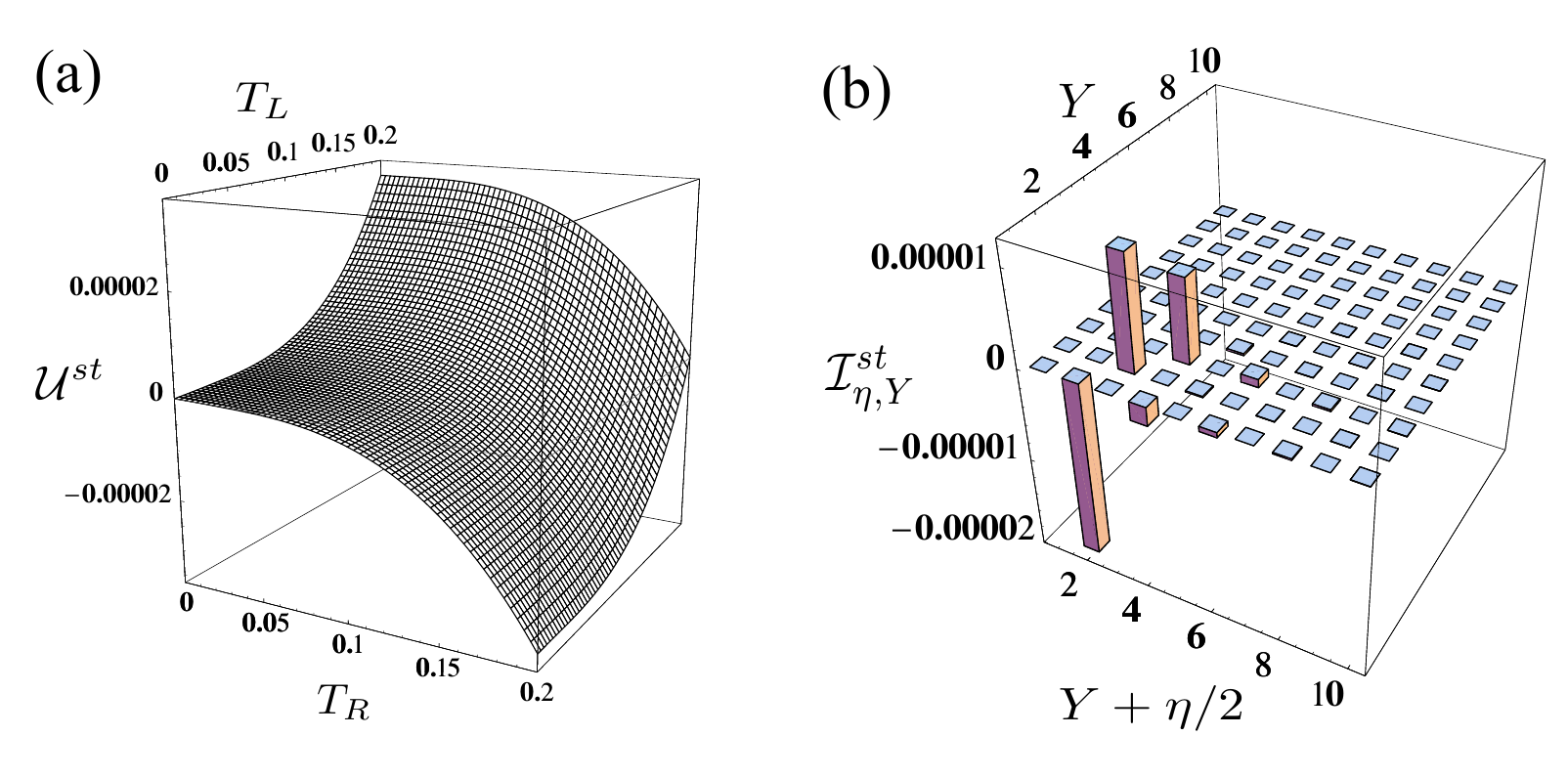}
\caption{The energy flow (a) and the particle current (b) of ome-dimensional base pairs of DNA molecule with 10 bases.
In (b) $T_{L}=0.07$ (350K) and $T_{R}=0.06$ (300K) are used.
 }
\label{Fig:DNA1} 
\end{center}
\end{figure}

As an application of our result, let us give an example of a real physical system, a hole transfer in one-dimensional chain of DNA base pairs.\cite{Berlin02}
In this case, a hole (or an electron) is an energy carrier and the molecular system has been described by the tight binding Hamiltonian given by Eq.(\ref{HM}).
It has been known that in a chain of DNA bases, the site energy of a guanine (G) and cytosine (C) base pair is higher than that of a thymine (T) and adenine (A) pair: $\varepsilon_{GC}-\varepsilon_{TA}\simeq 0.5$eV, and the transfer between the base pairs have been estimated to be $J\simeq 0.4$eV.
We show in Fig.\ref{Fig:DNA1}(a) a typical result of the energy flow when we put two GC pairs at the both ends of a molecular chain with its length of ${\cal N}=10$, where we have taken $g=0.5$.
It is found that the energy flow increases nonlinearly with temperature; Note that $T_{L,R}=300$K corresponds to $T_{L,R}=0.06$ in our unit.
The hole current in the molecular states is also shown in Fig.\ref{Fig:DNA1}(b) and the schematic picture of the energy flow and the particle current is shown in Fig.\ref{Fig:DNA1}(c), where we fix $T_{L}=0.07$ and $T_{R}=0.06$ corresponding to 350K and 300K, respectively.
It is found that the energy is transported by a particle current in the lower molecular states.

\section{concluding remarks}\label{Sec:Remarks}

We have obtained the nonequilibrium stationary state under thermal force as the zero eigenstate of the Liouvillian of a molecular chain which is weakly coupled with different thermal baths at the both ends.
The zero eigenstate is represented in the expansion of the order of correlation following the principle of the dynamics of correlation.
The physical quantity in the nonequilibrium stationary state is derived by taking an expectation value of an observable with respect to the stationary state.
The energy flow and the particle current are attributed to the first order and second order correlations created from the vacuum of correlation, respectively.
Since the first order correlation is directly related to the collision operator in kinetic theory, the energy flow can be cast into the form of Landauer equation with a transmission function with a strong resonance which reflects a discrete level structure of the molecular states.
It is also found that the thermal force induces an polarization, or its conjugate particle current, which cannot be cast into the Landauer type formula, because the particle current is attributed to the second order correlation.
Even so, there is a correspondence between them in their temperature dependences.

Here we have dealt with the nonequilibrium transport process of a molecule coupling with a thermal phonon bath with very broad spectral width.
It is interesting to investigate how the energy flow will be changed if we modify the phonon density of states in such a way that a particular molecular states be resonantly excited.
By replacing a phonon field with a radiation field in the present work, we can investigate how the energy flow behaves under a monochromatic light excitation in a nonequilibrium stationary state.
These further extensions of the present model will be shown elsewhere.

\acknowledgements

The authors thank Profs. N. Hatano and H. Hayakawa, and Mr. R. Tatsumi for fruitful discussions. This work was supported by the Grant-in-Aid for Scientific Research from the Ministry of Education, Science, Sports, and Culture of Japan and partially supported by Yukawa International Program for Quark-Hadron Sciences YIPQS.

\appendix

\section{Representation in Liouville space and the expression of $\mathcal{L}_{MB}$}
\label{App:LiouvilleSpace}

In this section, we shall briefly review the Liouville space representation of a Hilbert space operator.
The Liouville space is spanned by linear operators in $A, B, \cdots$ in the ordinary wave function space.\cite{Petrosky97}
As usual, the inner product of the Liouville space is defined by
\begin{equation}\label{App:InPrdct}
\<\!\< A|B\>\!\> = \Tr(A^{\dagger}B) \;,
\end{equation}
where $A$ and $B$ are linear operators acting on wave functions, and $A^{\dagger}$ is a Hermite conjugate of $A$.
For the case where the wave function space is spanned by a complete orthonormal basis,
\begin{equation}\label{App:Ortho}
\sum_{\alpha}|\alpha\>\<\alpha |=1\;, \<\alpha|\beta\> =\delta_{\alpha,\beta} \;,
\end{equation}
the Liouville space is spanned by a complete orthonormal basis of the dyads 
$
|\alpha ; \beta \>\!\>\equiv |\alpha \>\<\beta| \;,
$
i.e.,
\begin{equation}
\sum_{\alpha,\beta} |\alpha ; \beta \>\!\> \<\!\< \alpha ; \beta | =1 \; , \; \<\!\<\alpha ; \beta |\alpha' ; \beta' \>\!\>=\delta_{\alpha,\alpha'}\delta_{\beta,\beta'} \;.
\end{equation}
The matrix element of the usual operator $A$ in the wave function space is given by
\begin{equation}
\<\!\<\alpha ; \beta |A\>\!\>=\<\alpha|A|\beta\>  \;.
\end{equation}

The Liouville basis is constructed of the tensor product of the eigenstates of the unperturbed Hamiltonian.
For the molecular system and the thermal bath systems, the Liouville basis are written by
\begin{equation}
|\bar j_{1};\bar j_{2}\>\!\>\equiv |\bar j_{1}\>\<\bar j_{2}| \;,
\end{equation}
where $|\bar j\>$ denotes the eigenstate of $H_{M}$ given by Eq.(\ref{Barj}), 
and
\begin{equation}
|n_{r,{\bf q}};n'_{r,{\bf q}}\>\!\>\equiv |n_{r,{\bf q}}\>\<n'_{r,{\bf q}}| \;,
\end{equation}
where $|n_{r,{\bf q}}\>$ is the number state for a thermal bath normal mode $(r,{\bf q})$ of $H_{B}$.
In order to clarify the order of correlation, we prefer to use the Wigner basis defined by Eqs.(\ref {WignerPtcl}) and (\ref{WignerBath}):
\begin{equation}\label{App:WignerPtcl}
| \eta,Y \>\!\> \equiv |\bar j_{1};\bar j_{2}\>\!\>  \; ,
\end{equation}
where 
\begin{equation}
\eta\equiv \bar j_{1}-\bar j_{2} \;,\; Y\equiv {\bar j_{1}+\bar j_{2}\over 2} \;,
\end{equation}
and
\begin{equation}\label{App:WignerBath}
|\nu_{r,{\bf q}}, N_{r,{\bf q}}\>\!\>\equiv |n_{r,{\bf q}};n'_{r,{\bf q}} \>\!\> \;,
\end{equation}
where $\nu_{r,{\bf q}}$ and $N_{r,{\bf q}}$ are defined in Eq.(\ref{WignerBath}).
The Wigner basis is the eigenstate of ${\cal L}_{0}$ as shown in Eq.(\ref {L0EV}).
The Wigner basis then form the complete orthonormal basis satisfying
\begin{eqnarray}\label{App:WignerComp}
&&\<\!\< \eta,Y|\eta',Y'\>\!\>\!\otimes\<\!\<\{\nu\},\{N\}|\{\nu'\},\{N'\}\>\!\>=\delta_{\eta,\eta'}\delta_{Y,Y'} \nonumber\\
&& \qquad\qquad \cdot\delta_{\{\nu\},\{\nu'\}}\delta_{\{N\},\{N'\}}  \;,\\
 &&\sum_{\eta,Y}| \eta,Y\>\!\>\<\!\<\eta,Y|\sum_{\{\nu\},\{N\}}|\{\nu\},\{N\}\>\!\>\<\!\<\{\nu\};\{N\}|=1  \;, \nonumber\\
\end{eqnarray}
where $|\{\nu\},\{N\}\>\!\>\equiv \prod_{r,{\bf q}}|\nu_{r,{\bf q}},N_{r,{\bf q}}\>\!\>$.

\begin{widetext}

\begin{figure}[tp]
\begin{center}
\includegraphics[width=12cm]{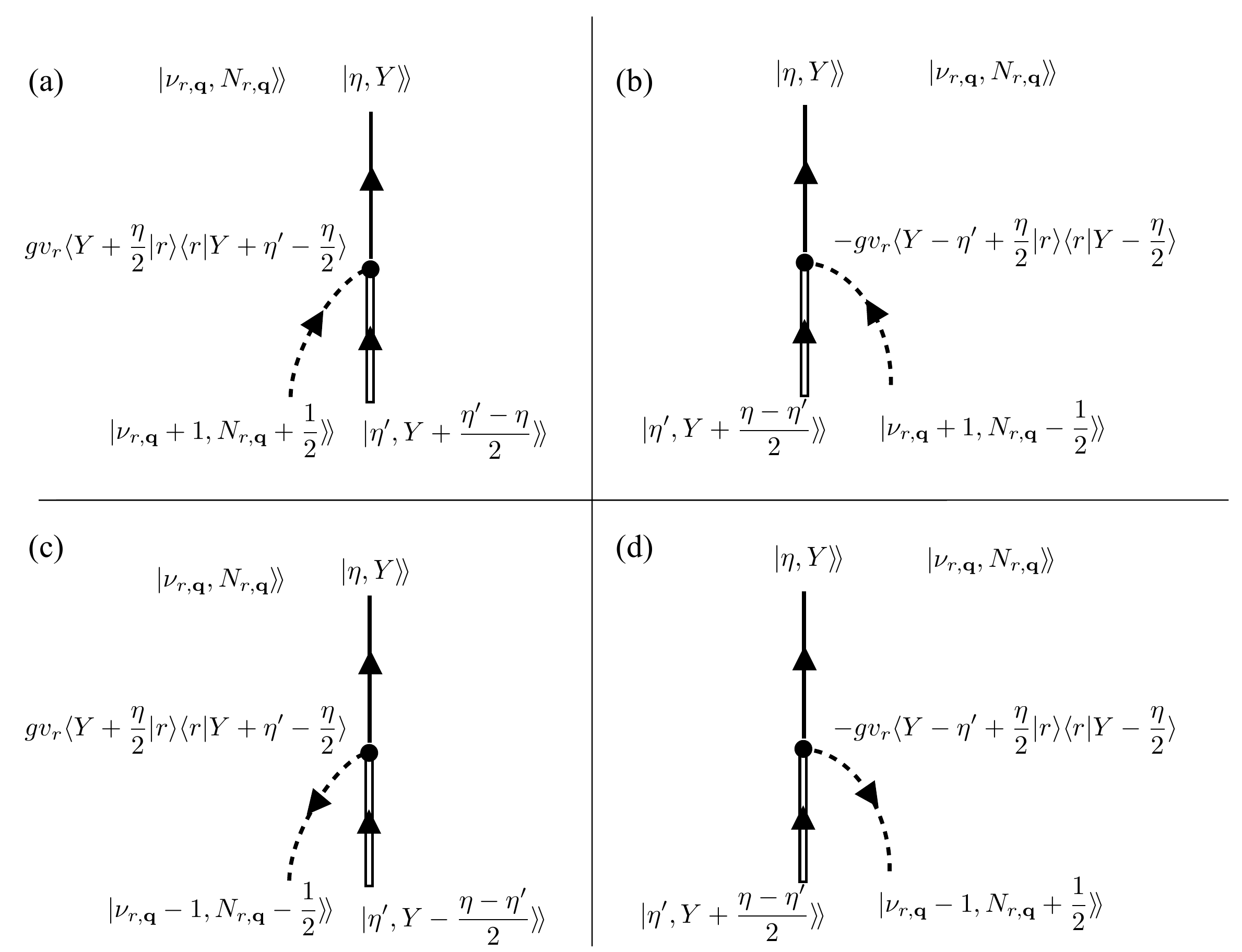}
\caption{The diagram of the interaction Liouvillian $L_{MB}$, where the lines represent the correlation.\cite{PrigogineText,Petrosky97}
The dotted line denotes the correlation, $\nu_{r,{\bf q}}$, of the phonon mode involved in the interaction, and the solid or double solid lines denote the correlation, $\eta$ or $\eta'$, of the molecular state.
The filled circle stands for the vertex of the interaction whose matrix element is written at the vertex. }
\label{Fig:LMBdiag} 
\end{center}
\end{figure}

\end{widetext}

Now we consider the matrix element of  the interaction Liouvillian ${\cal L}_{MB}\equiv H_{MB}\times 1 -1\times H_{MB}$ in terms of these Wigner basis.
The calculation of the matrix elements of ${\cal L}_{MB}$ can be done in a straightforward manner, yielding
\begin{widetext}
\begin{subequations}\label{App:LMBcomp}
\begin{eqnarray}
&&\<\!\<\eta,Y|\<\!\<\{\nu\};\{N\}|g{\cal L}_{MB}|\eta',Y'\>\!\>|\{\nu'\},\{N'\}\>\!\>  ={g\over\hbar\sqrt{\Omega}}\sum_{r=L,R}\sum_{\bf q} v_{r}\nonumber \\
&&\times\bigg[   \<Y+{\eta\over 2}|r\>\<r|Y'+{\eta'\over 2}\>\delta_{Y'-{\eta'\over2},Y-{\eta\over2}}\sqrt{N_{r,{\bf q}}+{\nu_{r,{\bf q}}\over2}+1} \;e^{{1\over2}{d\over dN_{r,{\bf q}}}}\delta_{\nu'_{r,{\bf q}},\nu_{r,{\bf q}}+1} \delta'_{\{\nu\},\{\nu'\}} \delta_{\{N\},\{N'\}} \\
&&  -  \<Y'-{\eta'\over 2}|r\>\<r|Y-{\eta\over 2}\>\delta_{Y'+{\eta'\over2},Y+{\eta\over2}}\sqrt{N_{r,{\bf q}}-{\nu_{r,{\bf q}}\over2}} \;e^{-{1\over2}{d\over dN_{r,{\bf q}}}}\delta_{\nu'_{r,{\bf q}},\nu_{r,{\bf q}}+1} \delta'_{\{\nu\},\{\nu'\}} \delta_{\{N\},\{N'\}}  \\
&&  +  \<Y'+{\eta'\over 2}|r\>\<r|Y+{\eta\over 2}\>\delta_{Y'-{\eta'\over2},Y-{\eta\over2}}\sqrt{N_{r,{\bf q}}+{\nu_{r,{\bf q}}\over2}} \;e^{-{1\over2}{d\over dN_{r,{\bf q}}}}\delta_{\nu'_{r,{\bf q}},\nu_{r,{\bf q}}-1} \delta'_{\{\nu\},\{\nu'\}} \delta_{\{N\},\{N'\}} \\
&&  -  \<Y-{\eta\over 2}|r\>\<r|Y'-{\eta'\over 2}\>\delta_{Y'+{\eta'\over2},Y+{\eta\over2}}\sqrt{N_{r,{\bf q}}-{\nu_{r,{\bf q}}\over2}+1} \;e^{{1\over2}{d\over dN_{r,{\bf q}}}}\delta_{\nu'_{r,{\bf q}},\nu_{r,{\bf q}}-1} \delta'_{\{\nu\},\{\nu'\}} \delta_{\{N\},\{N'\}} \bigg]
 \;,
\end{eqnarray}
\end{subequations}
\end{widetext}
where $\delta'$ stands for the product of the Kronecker delta except for the interaction normal mode of $(r,{\bf q})$.
We draw the diagram of the correlation corresponding to these four terms in Fig.\ref{Fig:LMBdiag}.
In the figures, the dotted line denotes the correlation, $\nu_{r,{\bf q}}$, of the phonon mode involved in the interaction, and the solid or double solid lines denote the correlation, $\eta$ or $\eta'$, of the molecular state.
The filled circle stands for the vertex of the interaction whose matrix element is written at the vertex.
The diagrams of (a) and (c) correspond to Eqs.(\ref {App:LMBcomp}a) and (\ref {App:LMBcomp}c), each of which is attributed to the  $b_{r,{\bf q}}$ (phonon absorption) and $b_{r,{\bf q}}^{\dagger}$ (phonon emission) terms in $H_{MB}\times 1$ term of ${\cal L}_{MB}$, where the phonon line appearing on the left side of the vertex implies that the transition of the particle ket state of $|Y'+\eta'/2\>$ to $|Y+\eta/2\>$ occurs by the interaction.
On the other hand, the diagrams of (b) and (d) correspond to Eqs.(\ref {App:LMBcomp}b) and (\ref {App:LMBcomp}d), each of which is attributed to the  $b_{r,{\bf q}}^{\dagger}$ (phonon emission) and $b_{r,{\bf q}}$ (phonon absorption) terms in $1\times H_{MB}$ term of ${\cal L}_{MB}$,  where the phonon line appearing on the right side of the vertex implies that the transition of the particle bra state of $\<Y'-\eta'/2|$ to $\<Y-\eta/2|$ occurs by the interaction.
The reader may be referred to the textbook about the way of writing the correlation lines.\cite{PrigogineText}

\section{The Complex Spectral Representation of Liouvillean and Subdynamics }
\label{App:Complex}

In this section, we shall summarize the complex spectral representation of Liouvillian.\cite{Petrosky97}
Useful formula for this paper are listed without proof.
The reader may refer to some references for detail.\cite{Petrosky97,Petrosky02,Petrosky10}

In the complex spectral representation of Liouvillian, we consider the eigenvalue problem for each correlation subspace $(\mu)=(\eta,\nu)$, where $\eta$ and $\nu$ represent the order of the correlation of the particle and phonon, respectively.

The complex eigenvalue problem is written as
 \begin{eqnarray}\label{App:EigenEq}
\mathcal{L}|F_{j}^{(\mu)} )\!) = Z_{j}^{(\mu)}|F_{j}^{(\mu)} )\!) \;, \; (\!( {\tilde F}_{j}^{(\mu)}|  \mathcal{L} =(\!( {\tilde F}_{j}^{(\mu)}| Z_{j}^{(\mu)} \;,
\end{eqnarray}
where the Liouvillian  can have complex eigenvalues $\mathrm{Im} Z_{j}^{(\mu)}\neq 0$.
It has been shown that the time evolution splits into two semigroups; one is oriented toward our future $t>0$ with $\mathrm{Im} Z_{j}^{(\mu)}<0$ (equilibrium is approached for $t\rightarrow \infty$), while the other is oriented toward our past $t<0$ with $\mathrm{Im} Z_{j}^{(\mu)}>0$.
All irreversible processes have the same time orientation.
To be self-consistent we choose the semigroup oriented toward our future, which determines the direction of the analytic continuation of the eigenfunction of $\mathcal{L}$.\cite{Petrosky96, Petrosky97}

Now we introduce the projection operators defined in (\ref{Pclass}) which satisfy 
\begin{subequations}
\begin{eqnarray}
&&\mathcal{L}_{0}\hat{\cal P}^{(\mu)}=\hat{\cal P}^{(\mu)}\mathcal{L}_{0} \; , \\
&& \hat{\cal P}^{(\mu)}\hat{\cal P}^{(\mu')}=\hat{\cal P}^{(\mu)}\delta_{\mu,\mu'} \;,  \\
&& \sum_{\mu}\hat{\cal P}^{(\mu)}=1 \;.
\end{eqnarray}
\end{subequations}
We also introduce the projection operators
\begin{equation}\label{App:Qk}
\hat{\cal Q}^{(\mu)}=1-\hat{\cal P}^{(\mu)}
\end{equation}
which are orthogonal to $\hat{\cal P}^{(\mu)}$.

We solve the eigenvalue problem (\ref{App:EigenEq}) for the perturbed system with $g\neq 0$ under the boundary conditions for the unperturbed case:
 \begin{eqnarray}
&&|F_{j}^{(\mu)} )\!)  =\hat{\cal P}^{(\mu)} |F_{j}^{(\mu)} )\!) \;,  \nonumber\\
&&(\!({\tilde F}_{j}^{(\mu)}| =(\!({\tilde F}_{j}^{(\mu)}| \hat{\cal P}^{(\mu)}  \quad{\mathrm{for}}\;  g=0 \;.
\end{eqnarray}
Hence, $\hat{\cal Q}^{(\mu)}|F_{j}^{(\mu)} )\!)=0$  for $g=0$.
The $\hat{\cal Q}^{(\mu)}$ components are created through the interaction for $g\neq 0$. 
The right and left eigenstates, $|F_{j}^{(\mu)} )\!)$ and $(\!({\tilde F}_{j}^{(\mu)}|$, are biorthonormal sets satisfying
\begin{equation}\label{App:BiOrthoF}
(\!(\tilde{F}_{j}^{(\mu)}|F_{j'}^{(\mu')})\!) =\delta_{j,j'}\delta_{\mu,\mu'}\; , \sum_{\mu,j}|F_{j}^{(\mu)})\!)(\!(\tilde{F}_{j}^{(\mu)}|=1  \; .
\end{equation}

Applying the projection operators $\hat{\cal P}^{(\mu)}$ and $\hat{\cal Q}^{(\mu)}$ in (\ref{Pclass}) and (\ref{App:Qk}) to the  (\ref{App:EigenEq}), we derive the set of equations:
\begin{subequations}\label{App:PQF}
\begin{eqnarray}   
&&\hat{\cal P}^{(\mu)}\mathcal{L}\Big(\hat{\cal P}^{(\mu)}|F_{j}^{(\mu)})\!)+\hat{\cal Q}^{(\mu)}|F_{j}^{(\mu)})\!) \Big)=Z_{j}^{(\mu)}\hat{P}^{(\mu)}|F_{j}^{(\mu)}{\>\!\>}   \;, \nonumber\\
&&\\
&&\hat{\cal Q}^{(\mu)}\mathcal{L}\Big(\hat{\cal P}^{(\mu)}|F_{j}^{(\mu)})\!)+\hat{\cal Q}^{(\mu)}|F_{j}^{(\mu)})\!) \Big)=Z_{j}^{(\mu)}\hat{\cal Q}^{(\mu)}|F_{j}^{(\mu)})\!)  \;.\nonumber\\
\end{eqnarray}
\end{subequations}

Equation (\ref{App:PQF}b) leads to
\begin{equation} \label{QF}
 \hat{\cal Q}^{(\mu)}|F_j^{(\mu)} )\!) =\hat{\mathcal{C}}^{(\mu)}(Z_j^{(\mu)}) \hat{\cal P}^{(\mu)} | F_j^{(\mu)})\!)   \;,
 \end{equation}
 where 
\begin{equation}
 \hat{\mathcal{C}}^{(\mu)}(z)=\frac{1}{z-\hat{\cal Q}^{(\mu)}\mathcal{L}\hat{\cal Q}^{(\mu)}}\hat{\cal Q}^{(\mu)}g\mathcal{L}_{MB}\hat{\cal P}^{(\mu)}
\end{equation}
is called the creation-of-correlation operator, or simply the {\it creation operator}.\cite{Petrosky97}
Substituting (\ref{QF}) into (\ref{App:PQF}a), we obtain
\begin{equation}\label{App:EVPCol2}
 \hat{\Psi}^{(\mu)} (Z_j^{(\mu)})|u_j^{(\mu)})\!) = Z_j^{(\mu)} | u_j^{(\mu)} )\!)   \; ,
\end{equation}
where 
\begin{equation}\label{App:uj2}
 |u_j^{(\mu)} )\!) = (N_j^{(\mu)})^{-1/2} \hat{\cal P}^{(\mu)} |F_j^{(\mu)} )\!) 
\end{equation}
and $N_j^{(\mu)}$ is a normalization constant which will be determined later.
Here, $\hat{\Psi}^{(\mu)}$ is the {\it collision operator} familiar to nonequilibrium statistical mechanics.\cite{PrigogineText, Resibois67, ResiboisBook,BalescuBook} 
This operator is associated to {\it diagonal transitions} between two states corresponding to the same projection operator $\hat{P}^{(\mu)}$:
\begin{eqnarray}
 \hat{\Psi}^{(\mu)}(z)&=&{\hat{\cal P}}^{(\mu)} \mathcal{L}_0 \hat{\cal P}^{(\mu)} +\hat{\cal P}^{(\mu)} g\mathcal{L}_{MB}\hat{\cal P}^{(\mu)} \nonumber\\ 
&+&\hat{\cal P}^{(\mu)}g\mathcal{L}_{MB}\hat{\cal Q}^{(\mu)}\hat{\mathcal{C}}^{(\mu)}(z) \hat{\cal P}^{(\mu)} \; .
\end{eqnarray}
Note that (\ref{App:EVPCol2}) is a nonlinear equation in the same sense of the Brillouin-Wigner perturbation method, i.e., the eigenvalue $Z_{j}^{(\mu)}$ appears in the collision operator.

Assuming completeness in the space $\hat{\cal P}^{(\mu)}$, we may always construct a set of states $\{(\!(\tilde{u}_{j}^{(\mu)}|\}$ biorthogonal to $\{|{u}_{j}^{(\mu)})\!)\}$, i.e.,
\begin{equation}\label{App:CompUj}
(\!(\tilde{u}_{j}^{(\mu)}|u_{j'}^{(\mu')})\!) =\delta_{j,j'}\delta_{\mu,\mu'}\; , \sum_{\mu,j}|u_{j}^{(\mu)})\!)(\!(\tilde{u}_{j}^{(\mu)}|=1  \; .
\end{equation}

The equation (\ref{App:EVPCol2}) combined with (\ref{App:uj2}) shows that the $\hat{\cal P}^{(\mu)}$ component of $|F_{j}^{(\mu)})\!)$ (which is called ``{\it privileged component}'' of $|F_{j}^{(\mu)})\!)$) is an eigenstates of the collision operator, which has the same eigenvalue $Z_{j}^{(\mu)}$ as the Liouvillean.
The solution of the eigenvalue problem of the Liouvillean for our class of singular functions has unique features.
The privileged components satisfy closed equations and the $\hat{\cal Q}^{(\mu)}$ components are ``driven'' by the privileged components [See (\ref{QF})].

Combining (\ref{QF}) with (\ref{App:uj}), we obtain the right eigenstates of the Liouvillean given by
\begin{equation}\label{App:Fright2}
|F_{j}^{(\mu)})\!)=\sqrt{N_{j}^{(\mu)}} \Big(\hat{\cal P}^{(\mu)}+\hat{\mathcal{C}}^{(\mu)}(Z_{j}^{(\mu)}) \Big) |u_{j}^{(\mu)})\!) \; .
\end{equation}
Similarly, we obtain for the left eigenstates given by 
\begin{equation}\label{App:Fleft}
(\!( \tilde{F}_{j}^{(\mu)}|=(\!( \tilde{v}_{j}^{(\mu)}| \Big(\hat{\cal P}^{(\mu)}+\hat{\mathcal{D}}^{(\mu)}(Z_{j}^{(\mu)}) \Big)\sqrt{N_{j}^{(\mu)}}  \; ,
\end{equation}
where the operator $\hat{\mathcal{D}}^{(\mu)}(Z_{j}^{(\mu)})$ is called the destruction-of-correlation operator, or the {\it destruction operator} for short, and is defined by [cf. (\ref{App:Creation})]
\begin{equation}\label{App:Destruction}
 \hat{\mathcal{D}}^{(\mu)}(z)=\hat{\cal P}^{(\mu)}g\mathcal{L}_{MB}\hat{\cal Q}^{(\mu)}\frac{1}{z-\hat{\cal Q}^{(\mu)}\mathcal{L}\hat{\cal Q}^{(\mu)}}\hat{\cal Q}^{(\mu)} \;.
\end{equation}
Again $ \hat{\mathcal{D}}^{(\mu)}(z)$ corresponds to the off-diagonal transitions; $ \hat{\mathcal{D}}^{(\mu)}(z)=\hat{\cal P}^{(\mu)} \hat{\mathcal{D}}^{(\mu)}(z)\hat{\cal Q}^{(\mu)}$.
Using $\hat{\mathcal{D}}^{(\mu)}(z)$, the collision operator $\hat{\Psi}^{(\mu)}(z)$ is also written as
\begin{eqnarray}\label{App:PsiUsingD}
 \hat{\Psi}^{(\mu)}(z)&=&{\hat{\cal P}}^{(\mu)} \mathcal{L}_0 \hat{\cal P}^{(\mu)} +\hat{\cal P}^{(\mu)} g\mathcal{L}_{MB}\hat{\cal P}^{(\mu)} \nonumber\\ 
&+&\hat{\cal P}^{(\mu)}\hat{\mathcal{D}}^{(\mu)}(z) \hat{\cal Q}^{(\mu)} g\mathcal{L}_{MB} \hat{\cal P}^{(\mu)} \; .
\end{eqnarray}

The normalization constant $N_{j}^{(\mu)}$ is determined by  Eq.(\ref{App:BiOrthoF})with  Eqs.(\ref{App:Fright2}) and (\ref{App:Fleft}) as
\begin{equation}\label{App:Nnorm}
\left( N_{j}^{(\mu)}\right)^{-1}=(\!(\tilde v_{j}^{(\mu)}|[\hat{\cal P}^{(\mu)}+\hat{\cal D}^{(\mu)}(Z_{j}^{(\mu)})\hat{\cal C}^{(\mu)}(Z_{j}^{(\mu)})]|u_j^{(\mu)})\!)\;.
\end{equation}

The states $(\!(\tilde{v}_{j}^{(\mu)}|$ are the left eigenstates of the collision operator $\hat{\Psi}^{(\mu)}$,
\begin{equation}\label{App:EVPColLeft}
(\!( \tilde{v}_{j}^{(\mu)}| \hat{\Psi}^{(\mu)}(Z_{j}^{(\mu)})=(\!( \tilde{v}_{j}^{(\mu)}| Z_{j}^{(\mu)} \;.
\end{equation}
We have revealed the correspondence between the eigenvalue problems of the Liouvillean $\mathcal{L}$ and the collision operator $\hat{\Psi}^{(\mu)}$.

\begin{widetext}

\section{Collision operator in the vacuum of correlation subspace}
\label{App:DerivKinEq}

In this section, we shall give the explicit expression of the collision operator of $\bar{\Psi}^{(0)}_{2}$  by applying Eqs.(\ref {App:LMBcomp}) into Eq.(\ref {Psi2Expression}).
We show the diagrams of $\bar\Psi_{2}^{(0)}(+i0)$ in Fig.\ref {Fig:CollisionDiag}.
The diagrams (a), (b), (c), and (d) represent the loss of the state $|0,Y\>\!\>$ which are written by
\begin{eqnarray}\label{App:Collision term loss}
&(a):&\<\!\<0,Y|\bar\Psi_2^{(0)}|0,Y\>\!\>_{(a)}={g^2\over \hbar^2}  {1\over \Omega}\sum_q \sum_{r=L,R}\sum_{\eta} |v_r|^2|c_{r,Y}^* c_{r,Y-\eta}|^2 {1\over i 0^+ - (\Delta_{-\eta,Y-\eta/2}+\omega_{r,q}) } \left(n_{r}(\omega_{r,q})+1\right)   \;,\\
&(a'):&\<\!\<0,Y|\bar\Psi_2^{(0)}|0,Y\>\!\>_{(a')}={g^2\over \hbar^2}  {1\over \Omega}\sum_q \sum_{r=L,R} \sum_{\eta} |v_r|^2|c_{r,Y}^* c_{r,Y+\eta}|^2 {1\over i 0^+ - (\Delta_{\eta,Y+\eta/2}-\omega_{r,q}) } n_{r} (\omega_{r,q})   \;,\\
&(b):&\<\!\<0,Y|\bar\Psi_2^{(0)}|0,Y\>\!\>_{(b)}={g^2\over \hbar^2}  {1\over \Omega}\sum_q \sum_{r=L,R}\sum_{\eta} |v_r|^2|c_{r,Y}^* c_{r,Y+\eta}|^2 {1\over i 0^+ - (\Delta_{-\eta,Y+\eta/2}+\omega_{r,q}) } n_{r} (\omega_{r,q})   \;,\\
&(b'):&\<\!\<0,Y|\bar\Psi_2^{(0)}|0,Y\>\!\>_{(b')}={g^2\over \hbar^2} {1\over \Omega}\sum_q  \sum_{r=L,R}\sum_{\eta} |v_r|^2|c_{r,Y}^* c_{r,Y-\eta}|^2 {1\over i 0^+ - (\Delta_{\eta,Y-\eta/2}-\omega_{r,q}) }\left( n_{r} (\omega_{r,q})+1 \right)    \;.
\end{eqnarray}
where $n_{r}(\omega)$ is Planck's distribution function given by Eq.(\ref{Planck}).

\begin{figure}[b]
\begin{center}
\includegraphics[width=12cm,height=10cm]{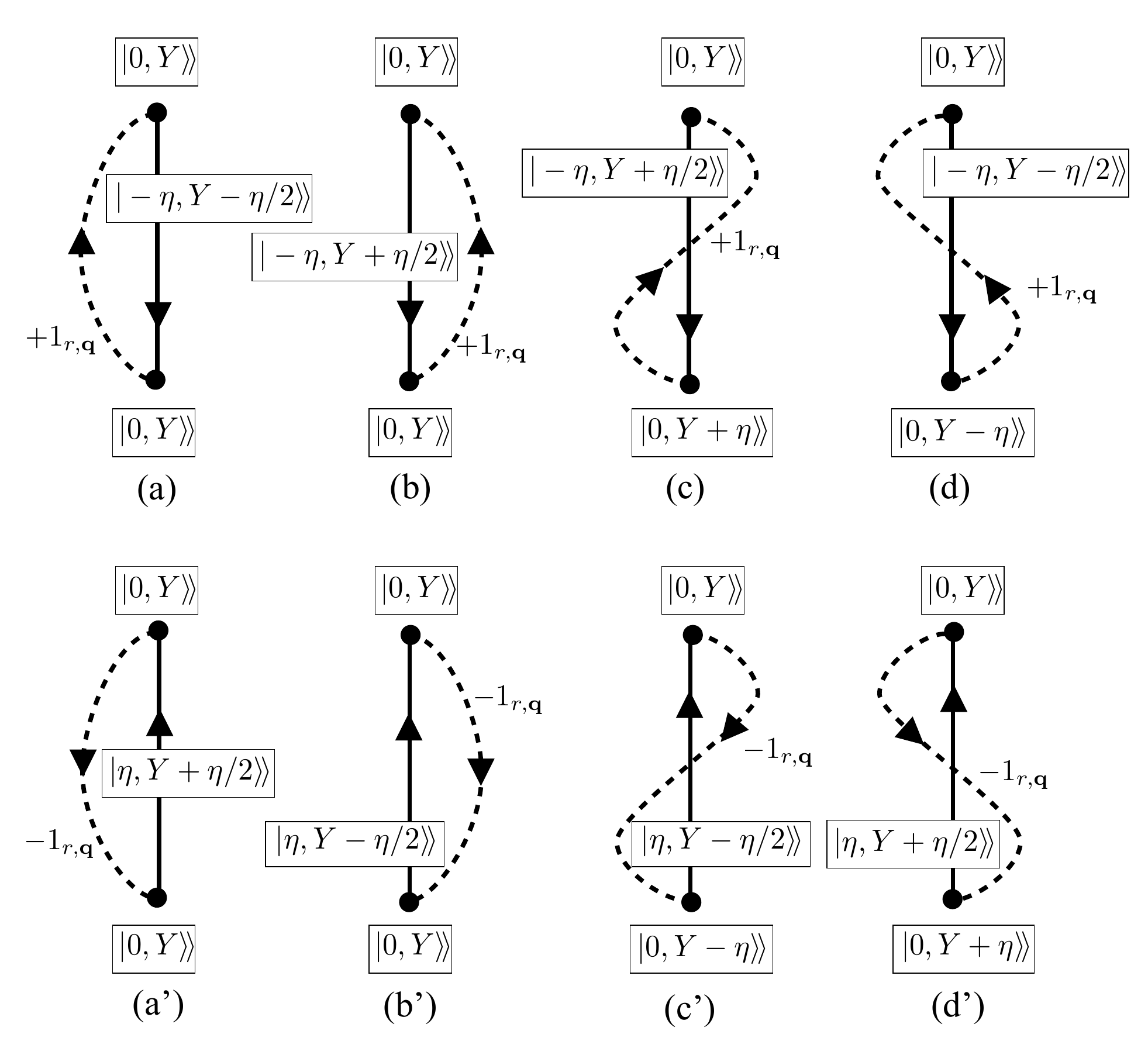}
\caption{ Diagram of the collision operator $\overline{\Psi}_{2}^{(k)}(+i0)$. }
\label{Fig:CollisionDiag} \end{center}
\end{figure}

On the other hand, the diagrams (c), (c'), (d), and (d') represent the gain of the $|0,Y\>\!\>$ state from the $|0,Y\pm \eta\>\!\>$ states which are written as 
\begin{eqnarray}\label{App:Collision term gain}
&(c):&\<\!\<0,Y|\bar\Psi_2^{(0)}|0,Y+\eta\>\!\>_{(c)}=-{g^2\over \hbar^2}  {1\over \Omega}\sum_q  \sum_{r=L,R} |v_r|^2|c_{r,Y}^* c_{r,Y+\eta}|^2 {1\over i 0^+ - (\Delta_{-\eta,Y+\eta/2}+\omega_{r,q}) } \left(n_{r}(\omega_{r,q})+1\right)   \;,\\
&(c'):&\<\!\<0,Y|\bar\Psi_2^{(0)}|0,Y-\eta\>\!\>_{(c')}=-{g^2\over \hbar^2} {1\over \Omega}\sum_q  \sum_{r=L,R}  |v_r|^2|c_{r,Y}^* c_{r,Y-\eta}|^2 {1\over i 0^+ - (\Delta_{\eta,Y-\eta/2}-\omega_{r,q}) } n_{r}(\omega_{r,q})   \;,\\
&(d):&\<\!\<0,Y|\bar\Psi_2^{(0)}|0,Y-\eta\>\!\>_{(d)}=-{g^2\over \hbar^2}  {1\over \Omega}\sum_q \sum_{r=L,R} |v_r|^2|c_{r,Y}^* c_{r,Y-\eta}|^2 {1\over i 0^+ - (\Delta_{-\eta,Y-\eta/2}+\omega_{r,q}) } n_{r}(\omega_{r,q})   \;,\\
&(d'):&\<\!\<0,Y|\bar\Psi_2^{(0)}|0,Y+\eta\>\!\>_{(d')}=-{g^2\over \hbar^2} {1\over \Omega}\sum_q   \sum_{r=L,R} |v_r|^2|c_{r,Y}^* c_{r,Y+\eta}|^2 {1\over i 0^+ - (\Delta_{\eta,Y+\eta/2}-\omega_{r,q}) }\left(  n_{r}(\omega_{r,q}) +1 \right)   \;, 
\end{eqnarray}
where $n_{r}(\omega)$ is the Planck's distribution given by Eq.(\ref{Planck}).
The summation for the thermal phonon modes $q$ is replaced by the integral with use of the density of states given by Eq.(\ref {phonon DOS}).

When we sum up the contributions of these diagrams, we have  for $\eta>0$
\begin{eqnarray}
\<\!\<0,Y|\bar\Psi_2^{(0)}|0,Y\>\!\>&=&-2\pi i {g^2\over \hbar^2} \sum_{r=L,R}\sum_{\eta >0}  |v_r|^2 \Big\{
|c_{r,Y}^* c_{r,Y-\eta}|^2  {\cal D}_{ph}(\Delta_{\eta,Y-\eta/2})  (n_{r}(\Delta_{\eta,Y-\eta/2})+1 )  \nonumber\\
&& + |c_{r,Y}^* c_{r,Y+\eta}|^2 {\cal D}_{ph}( \Delta_{\eta,Y+\eta/2})  n_{r} ( \Delta_{\eta,Y+\eta/2})  \Big\}  \;, \\
\<\!\<0,Y|\bar\Psi_2^{(0)}|0,Y+\eta\>\!\>&=&2\pi i {g^2\over \hbar^2}   \sum_{r=L,R} |v_r|^2|c_{r,Y}^* c_{r,Y+\eta}|^2 {\cal D}_{ph}(\Delta_{\eta,Y+\eta/2}) \left(n_{r}(\Delta_{\eta,Y+\eta/2})+1\right)   \;,\\
\<\!\<0,Y|\bar\Psi_2^{(0)}|0,Y-\eta\>\!\>&=&2\pi i{g^2\over \hbar^2}  \sum_{r=L,R}  |v_r|^2|c_{r,Y}^* c_{r,Y-\eta}|^2 {\cal D}_{ph}(\Delta_{\eta,Y-\eta/2})  n_{r}(\Delta_{\eta,Y-\eta/2})   \;,
\end{eqnarray}
where we have used Eqs.(\ref{IntDOS}) and (\ref{PropDelta}).
When we define the transition probabilities as
\begin{eqnarray}\label{App:tranProb1}
k_{r}^{Y,Y+\eta}&\equiv& 2\pi  {g^2\over \hbar^2}   |v_r|^2|c_{r,Y}^* c_{r,Y+\eta}|^2 {\cal D}_{ph}(\Delta_{\eta,Y+\eta/2}) \left(n_{r}(\Delta_{\eta,Y+\eta/2})+1\right)   \;,\\
k_{r}^{Y,Y-\eta}&\equiv &2\pi {g^2\over \hbar^2}   |v_r|^2|c_{r,Y}^* c_{r,Y-\eta}|^2 {\cal D}_{ph}(\Delta_{\eta,Y-\eta/2})  n_{r}(\Delta_{\eta,Y-\eta/2})  \;,
\end{eqnarray}
where $\eta >0$, 
we can write
\begin{subequations}\label{App:tranProb2}
\begin{eqnarray}
&&\<\!\<0,Y|\bar\Psi_2^{(0)}|0,Y\>\!\>=-i\sum_{r=L,R}\sum_{\eta >0} \left(k_{r}^{Y-\eta,Y} +k_{r}^{Y+\eta,Y} \right)  \;,  \\
&&\<\!\<0,Y|\bar\Psi_2^{(0)}|0,Y+\eta\>\!\>= i\sum_{r=L,R} k_{r}^{Y,Y+\eta} \;,\\
&&\<\!\<0,Y|\bar\Psi_2^{(0)}|0,Y-\eta\>\!\>= i\sum_{r=L,R} k_{r}^{Y,Y-\eta}  \;.
\end{eqnarray}
\end{subequations}

\end{widetext}

For a ${\cal N}=2$ molecule, the eigenstates of $H_{M}$ are obtained as
\begin{subequations}\label{App:EigSt2}
\begin{eqnarray}
&&|\bar 1\>=c_{L,\bar 1}|L\>+c_{R,\bar 1}|R\>  \;,\\
&&|\bar 2\>=c_{L,\bar 2}|L\>+c_{R,\bar 2}|R\>  \;,
\end{eqnarray}
\end{subequations}
with
\begin{subequations}
\begin{eqnarray}
&&c_{L,\bar 1}=\cos \theta \;,\; c_{R,\bar 1}=\sin \theta \;,\\
&&c_{L,\bar 2}=-\sin \theta \;,\; c_{R,\bar 2}=\cos \theta \;,
\end{eqnarray}
\end{subequations}
where $\tan 2\theta = 2J/(E_{R}-E_{L})$.
The eigenvalues of these eigenstates are given by
\begin{subequations}
\begin{eqnarray}
&&E_{\bar 1}={ \varepsilon_{L}+\varepsilon_{R}-\sqrt{(\varepsilon_{L}-\varepsilon_{R})^{2}+4J^{2}}\over 2} \;,\\
&&E_{\bar 2}={ \varepsilon_{L}+\varepsilon_{R}+\sqrt{(\varepsilon_{L}-\varepsilon_{R})^{2}+4J^{2}} \over 2}\;.
\end{eqnarray}
\end{subequations}
The transition probabilities are then given by
\begin{eqnarray}
k_{L}^{1,2}&=& {\pi  g^2 v_{L}^2 \over 2\hbar^2} \sin^2 2\theta \;{\cal D}_{ph}(\Delta) \left(n_{L}(\Delta)+1\right)   \;,\\
k_{R}^{1,2}&=& {\pi  g^2 v_{R}^2 \over 2\hbar^2} \sin^2 2\theta \;{\cal D}_{ph}(\Delta) \left(n_{R}(\Delta)+1\right)   \;,\\
k_{L}^{2,1}&=& {\pi  g^2 v_{L}^2 \over 2\hbar^2} \sin^2 2\theta \;{\cal D}_{ph}(\Delta)  n_{L}(\Delta)   \;,\\
k_{R}^{2,1}&=& {\pi  g^2 v_{R}^2 \over 2\hbar^2} \sin^2 2\theta \;{\cal D}_{ph}(\Delta)  n_{R}(\Delta)   \;.
\end{eqnarray}

\section{Energy Flow in terms of a energy change of the thermal bath}
\label{App:BathChange}

In this section, we shall show that the formula of the energy flow Eq.(\ref {WstExp}) is also obtained by investigating the energy change of the thermal bath systems.
We can denote the energy flow from the left thermal bath as a energy change of the left thermal bath:
\begin{equation}\label{App:wLdef}
w^L(t)={d\over dt}\<H_B^L\>_t=\sum_{\bf q}\omega_{L,{\bf q}}{\rm Tr}[\hat N_{L,{\bf q}}{\cal L}_{MB}\rho(t) ] \;,
\end{equation}
where $\hat N_{L,{\bf q}}\equiv b_{L,{\bf q}}^\dagger b_{L,{\bf q}}$.
This definition coincides with the ordinary definition of the particle current from a particle bath to a mesoscopic system.\cite{Comment}
Using Eqs.(\ref {App:LMBcomp}) and replacing $\rho(t)$ with $F_{0}^{(0)}$, we have the expression of the energy flow from the left bath in the nonequilibrium stationary state as
\begin{equation}\label{App:wLst}
w_L^{st}={g v_L\over\sqrt{\Omega}}\sum_{\bf q}\sum_{\eta,Y}c_{L,Y+\eta}c_{L,Y}\rm{Tr}\big[(b_{L,{\bf q}}^\dagger-b_{L,{\bf q}})F_0^{(0)}\big] \;.
\end{equation}
Substituting Eq.(\ref{F00-2}) into Eq.(\ref{App:wLst}) and after some calculation, we obtain 
\begin{eqnarray}
w_L^{st}&=&\frac{2\pi g^2 v_L^2}{\hbar} \sum_{\eta>0}{\cal D}^{ph}(\Delta_{\eta,Y+\eta/2})\big|c_{L,Y+\eta}^{*}c_{L,Y}\big|^2  \Delta_{\eta,Y+\eta/2} \nonumber\\
&\times&\bigg\{ \phi_{Y}n_L(\Delta_{\eta,Y+\eta/2})-\phi_{Y+\eta} \big( n_L(\Delta_{\eta,Y+\eta/2})+1 \big) \bigg\} \;. \nonumber\\
\end{eqnarray}
The energy flow going out to the right thermal bath has been calculated similarly.
By summing up the two contributions to obtain the same results as given in Eq.(\ref{WstExp}).
These derivations mentioned here is to clarify that the energy flow which is carried by a phonon particle flowing from the left bath to the right one.
This is the reason why the energy flow in the present case is described by the Landauer formula which is mostly used to represent the electronic current in which an electron flows in from an electron reservoir at one end and going out to the other end through a one-dimensional mesescopic system.


\end{document}